\documentclass[a4paper,10pt]{article}
\usepackage[utf8]{inputenc}
\usepackage{authblk}

\usepackage{bm,amsmath,amssymb,natbib,graphicx}

\newcommand{\al}{\alpha}
\newcommand{\be}{\beta}
\newcommand{\ga}{\gamma}
\newcommand{\dd}{\mathrm{d}}
\newcommand{\fzero}{f^{(0)}}
\newcommand{\fone}{f^{(1)}}

\newcommand{\fb}{\bar{f}}

\newcommand{\ux}{\bm{x}}

\newcommand{\uxi}{\bm{\xi}}
\newcommand{\ua}{\bm{a}}
\newcommand{\ue}{\bm{e}}

\newcommand{\uN}{\bm{\nabla}}

\newcommand{\uk}{\bm{k}}

\newcommand{\uc}{\bm{c}}

\newcommand{\uP}{\bm{P}}
\newcommand{\uA}{\bm{A}}
\newcommand{\uS}{\bm{S}}

\renewcommand{\H}{\mathcal{H}}

\renewcommand{\Re}{\mathrm{Re}}
\renewcommand{\Im}{\mathrm{Im}}
\newcommand{\uI}{\bm{I}}
\newcommand{\uLambda}{\bm{\Lambda}}
\newcommand{\uu}{\bm{u}}

\newcommand{\p}{\partial}

\renewcommand{\O}{\mathcal{O}}
\newcommand{\Ma}{\mathrm{Ma}}

\title{Increasing stability and accuracy of the lattice Boltzmann scheme: recursivity and regularization}
\author[a,b]{Orestis Malaspinas\thanks{{orestis.malaspinas@unige.ch}}}

\affil[a]{Institut Jean le Rond d'Alembert, UMR 7190, Universit\'e Pierre et Marie Curie - Paris 6, 4 place Jussieu - case 
162, F-75252, France}
\affil[b]{Centre Universitaire d'Informatique, Universit\'e de Gen\`eve
7, route de Drize, CH-1227 Switzerland}

\date{\today}

\begin{document}

\maketitle
\begin{abstract}
 In the present paper a lattice Boltzmann scheme is presented
 which exhibits an increased stability and accuracy with respect to standard
 single- or multi-relaxa-tion-time (MRT) approaches. The scheme is based on 
 a single-relaxation-time model where a special regularization procedure
 is applied. This regularization is based on the fact that, for a-thermal flows, there exists a recursive
 way to express the velocity distribution function at any order (in the Hermite series sense)
 in terms of the density, velocity, and stress tensor. A linear stability analysis 
 is conducted which shows enhanced dispersion/dissipation relations with respect to 
 existing models. The model is then validated
 on two (one 2D and one 3D) moderately high Reynolds number simulations ($\Re\sim 1000$) at moderate Mach numbers
 ($\Ma\sim 0.5$). In both cases the results are compared with an MRT model and 
 an enhanced accuracy and stability is shown by the present model.
\end{abstract}

\section{Introduction}

The lattice Boltzmann method (LBM) is a widely used tool for numerical simulations of fluid flows.
It has become over the years one the of the major engineering tools 
for computational fluid mechanics. It describes the flow thanks to the time evolution
of the velocity distribution function which is only modified through the effect of inter-particle collisions.

The most commonly used lattice Boltzmann collision model is the single relaxation time model 
or BGK (for Bhathagar, Gross and Krook, see \cite{bi_bhatnagar54}). This model 
is able to asymptotically represent weakly compressible fluids
(through a Chapman--Enskog expansion, see \citet{bi_chapman60}).
Nevertheless it suffers from stability issues especially at high Reynolds numbers.
These issues are due to 
the ``ghost-modes'' (see \citet{bi_dellar01} for a discussion) which are non-physical 
moments present in any LBM simulation in excess of the density (pressure), velocity and stress.
This issue has been addressed by several authors and several solutions have been proposed. The first is the multiple-relaxation-time (MRT)
approach (see \citet{bi_dhumieres92, bi_lallemand00, bi_dhumieres02, bi_dellar01, bi_xu11, bi_xu12} among others)
which uses a more complex collision model involving several relaxation times
adjusted with the help of a linear stability analysis in order to optimize their 
dispersion/dissipation relations. The entropic approach 
ensures the positivity of the distribution functions and hence the unconditional stability by adding an $H$-theorem to the BGK model. 
The effect of the $H$-theorem is essentially to increase locally the viscous dissipation
of the model (see \citet{bi_ansumali02,bi_boghosian03,bi_chikatamarla06,bi_malaspinas08} among others). 
The regularization approach (see \citet{bi_latt06,bi_zhang06}) which can be
interpreted as a subclass of MRT methods where the ``ghost-modes'' are relaxed towards zero with 
characteristic time 1. Finally the selective viscosity models proposed by \citet{bi_ricot09} 
use non-local low-pass filters to remove high frequency oscillations (which are responsible for the numerical instabilities)
in order to increase the stability.

In this paper we will first show a recursive way to compute the moments of the distribution
function as long as the Chapman--Enskog expansion is valid (low Knudsen number) for the BGK collision operator. 
This recursive relation
will then be used to ``regularize'' the distribution function and provide a very stable and accurate 
scheme even at moderately high Reynolds numbers and (relatively) high Mach numbers (smaller than one though).
We also show that the present model is more accurate and more stable than the existing MRT methods
by performing a linear stability analysis and several numerical benchmarks.

The paper is structured as follows. In Sec.~\ref{sec_lbm} a reminder of 
fundamentals for fluid flows with the Boltzmann--BGK equation is presented. Then in Sec.~\ref{sec_rec}
the new model is proposed and analyzed. It is validated in Sec.~\ref{sec_bench} on a 2D and a 3D benchmark.
Finally the present work is concluded in Sec.~\ref{sec_concl} and perspectives are given.

\section{The hydrodynamic limit of the BGK equation}\label{sec_lbm}

The following section aims at introducing the basic notations as well as showing the
fundamentals of the expansion leading from the continuous Boltzmann-BGK equation to 
the Navier--Stokes equations. More details can be found in \citet{bi_shan06} and \citet{bi_malaspinas-thesis} for example.

The Boltzmann equation describes the time evolution of the velocity density probability distribution $f(\ux,\uxi,t)$ of
finding a particle with velocity $\uxi$ at position $\ux$ and time $t$ in terms of particle collisions only, and
reads in absence of a force as
\begin{equation}
\p_t f(\ux,\uxi,t)+(\uxi\cdot\uN)f(\ux,\uxi,t)=\Omega(f),
\end{equation}
where $\Omega$ is the collision operator. Assuming also that the fluid is athermal (absence of temperature), the macroscopic fields of
interest, the 
density $\rho$, the velocity $\uu$, and the stress tensor $\uP$ are given by the following moments of the distribution function
\begin{align}
\rho&=\int\dd\uxi\  f(\ux,\uxi,t),\label{eq_def_rho}\\
\rho\uu&=\int\dd\uxi\  \uxi f(\ux,\uxi,t),\label{eq_def_j}\\
\uP&=\int\dd\uc\ \uc\uc f(\ux,\uxi,t),\label{eq_def_p}
\end{align}
where $\uc=\uxi-\uu$ is the microscopic velocity in the co-moving frame and $\uc\uc$ denotes the tensor product of $\uc$ with itself. 

The most widely used model for computational
fluid dynamics for the collision operator is the BGK,
single relaxation time approximation, in which the Boltzmann equation reads
\begin{equation}
\p_t f(\ux,\uxi,t)+(\uxi\cdot\uN)f(\ux,\uxi,t)=-\frac{1}{\tau}\left(f(\ux,\uxi,t)-\fzero(\rho,\uu)\right),\label{eq_bgk}
\end{equation}
where $\tau$ the relaxation time, and $\fzero$ is the local
Maxwell--Boltzmann equilibrium distribution function, which in non-dimensional units is given by 
\begin{equation}
\fzero=\frac{\rho(\ux,t)}{(2\pi)^{D/2}}
\exp\left(-\frac{(\uu(\ux,t)-\uxi)^2}{2}\right),
\end{equation}
$D$ being the physical dimension.

Since we are interested
in numerically solving Eq.~\eqref{eq_bgk} in an efficient fashion that nevertheless
represents accurately fluid flows, certain simplifications will be made. 
In particular instead of considering the complete form of the Maxwell--Boltzmann distribution
function, only a polynomial approximation will be used.

Following an idea by \citet{bi_shan06} (or \citet{bi_grad49b} for the original use of this expansion in the frame 
of the Boltzmann equation), 
one can expand the distribution functions $f$ and $\fzero$,
in Hermite polynomials up to an arbitrary order $N$ (see \citet{bi_grad49a} for a summary on Hermite polynomials)
\begin{align}
f^N=w(\uxi)\sum_{n=0}^N \frac{1}{n!}\H^{(n)}(\uxi):\ua^{(n)},\quad{\fzero}^N=w(\uxi)\sum_{n=0}^N \frac{1}{n!}\H^{(n)}(\uxi):\ua_0^{(n)},\label{eq_hermite_eq}
\end{align}
where the colon symbol ``:'' stands for the full index contraction. 
The Hermite polynomials of order $n$ and the associated Gaussian weight are noted $\H^{(n)}$ and $w(\uxi)=\exp(-\uxi^2/2)$ respectively.
The Hermite coefficients of $f$ and $\fzero$ of degree $n$ are respectively given by
$\ua^{(n)}$ and $\ua^{(n)}_0$.
From now on, we will always omit the superscript $N$ and assume that the distribution function (and its equilibrium counterpart) is represented 
by its approximate form in terms of Hermite polynomials up to an arbitrary order $N$
except when explicitly stated otherwise. The equilibrium coefficients can be easily computed and are found to be up to order three
\begin{align}
a_0^{(0)}&=\rho,\label{eq_a^0_0}\\
a_{0\al}^{(1)}&=\rho u_\al,\label{eq_a^1_0}\\
a_{0\al\be}^{(2)}&=\rho u_\al u_\be,\label{eq_a^2_0}\\
a_{0\al\be\ga}^{(3)}&=\rho u_\al u_\be u_\ga.\label{eq_a^3_0}
\end{align}

In order to recover the macroscopic equations of motion related with the BGK equation, one must
take moments of Eq.~\eqref{eq_bgk}. By taking the moments related with density (order zero) and momentum (order one) of this equation, 
one gets after some algebra and the use of Eqs.~\eqref{eq_def_rho}-\eqref{eq_def_p} and \eqref{eq_a^0_0}-\eqref{eq_a^2_0}
\begin{align}
&\p_t\rho+\uN\cdot(\rho\uu)=0,\\
&\p_t (\rho\uu)+\uN\cdot(\rho\uu\uu)+\uN\cdot\uP=0.
\end{align}
These equations are obtained under the sole assumption of mass and momentum conservation
($\int\Omega=\int\uxi\Omega=0$), or in other terms
\begin{align}
&\int\dd\uxi~(f-\fzero)=0,\quad\quad &\hbox{Mass conservation}\\
&\int\dd\uxi~\uxi (f-\fzero)=0,\quad &\hbox{Momentum conservation}
\end{align}

The momentum conservation equation still needs to be closed (a constitutive equation must be found for $\uP$). 
In order to do so, one can use the Chapman--Enskog
expansion (see \citet{bi_chapman60,bi_huang87}). Since the expansion in Hermite series 
is used for discretization purposes we will discuss the Chapman--Enskog expansion 
in this frame (although the Hermite series is not a prerequisite for performing the Chapman--Enskog expansion).

The Chapman--Enskog expansion is based on the assumption that the distribution 
function $f$ is given by the sum of the equilibrium distribution, $\fzero$, plus a small perturbation noted $\fone$,
\begin{equation}
f=\fzero+\fone,\label{eq_f-ce}
\end{equation}
where the equilibrium distribution is assumed to be given by Eq.~\eqref{eq_hermite_eq}. The perturbation, $\fone\sim \O(Kn)\ll \fzero$, is of the order of the Knudsen number, $Kn$. 
As for $f$ and $\fzero$ one can express $\fone$ in terms of a Hermite series
\begin{equation}
 \fone=w(\uxi)\sum_{n=0}^N \frac{1}{n!}\H^{(n)}(\uxi):\ua_1^{(n)},
\end{equation}
where $\ua_1^{(n)}$ is the Hermite coefficient of $\fone$ at order $n$.
The derivation which is presented hereafter is not the standard one found in the literature
and rather follows \citet{bi_huang87}.

Replacing the Chapman--Enskog Ansatz in Eq.~\eqref{eq_bgk}, one obtains at the lowest order
\begin{equation}
  \p_t\fzero+(\uxi\cdot\uN)\fzero=-\frac{1}{\tau}\fone.\label{eq_ce-bgk}
\end{equation}
Taking the zeroth and first order moments of this equation 
and using the mass and momentum conservation constrains on each
equation respectively ($\int\fone=\int\uxi\fone=0$),
one gets the inviscid Euler equations for mass, momentum and energy conservation
\begin{align}
&\p_t\rho+\uN\cdot(\rho\uu)=0,\label{eq_rho_euler}\\
&\p_t (\rho\uu)+\uN\cdot(\rho\uu\uu)+\uN p=0,\label{eq_j_euler}
\end{align}
where $p=\rho$ is the perfect gas law (remember that there is no temperature).

The stress tensor can be decomposed in its Chapman--Enskog counterparts
\begin{equation}
  \uP=\uP^{(0)}+\uP^{(1)}=\rho\uI+\uP^{(1)},
\end{equation}
where $\uP^{(j)}=\int\uc\uc f^{(j)}$ for $j=0,1$ ($j$ corresponding to the Chapman--Enskog index).
Thus we are left with the computation of $\uP^{(1)}$ which for simplicity
is computed through the Hermite expansion of the distribution function. Let us define
$\ua^{(n)}_1$ the Hermite coefficient of order $n$ of the off-equilibrium distribution function $\fone$
and express $\uP^{(1)}$ in terms of these Hermite coefficients 
\begin{equation}
P^{(1)}_{\al\be}=a^{(2)}_{1\al\be},
\end{equation}
where we used that by construction $\ua^{(0)}_1=\ua_1^{(1)}=0$.
Then projecting Eq.~\eqref{eq_ce-bgk} on the Hermite basis, it follows that
\begin{equation}
\p_t \ua_0^{(n)}+\uN\cdot \ua^{(n+1)}_0+\left(\uN \ua^{(n-1)}_0+\hbox{perm}\right)=-\frac{1}{\tau}\ua_1^{(n)},\label{eq_a0-a1-classical}
\end{equation}
where ``perm'' stands for all the cyclic index permutations. For $n=2$ this equation becomes
\begin{align}
&\p_t \ua_0^{(2)}+\uN\cdot \ua^{(3)}_0+\left(\uN \ua^{(1)}_0+\hbox{perm}\right)=-\frac{1}{\tau}\ua_1^{(2)},\nonumber\\
&\p_t \left(\rho\uu\uu\right)+\uN\cdot (\rho\uu\uu\uu)+\left(\uN (\rho\uu)+(\uN(\rho\uu))^\mathrm{T}\right)=-\frac{1}{\tau}\ua_1^{(2)}.
\end{align}
By using Eqs.~\eqref{eq_rho_euler}-\eqref{eq_j_euler} to eliminate the time derivative terms, this equation can
be rewritten (after some tedious algebra that can be found in \citet{bi_malaspinas-thesis}) as
\begin{equation}
\ua_1^{(2)}=\uP^{(1)}=-2\tau\rho\uS,\label{eq_a1_2-classical}
\end{equation}
where 
\begin{equation}
\uS=\frac{1}{2}\left(\uN\uu+(\uN\uu)^\mathrm{T}\right).\label{eq_lambda_P}
\end{equation}

By comparing Eq.~\eqref{eq_a1_2-classical} with the Navier--Stokes equations, the transport coefficient $\mu$
can be identified with
the relaxation time through the following relation
\begin{equation}
\mu=\rho\tau.\label{eq_nu_def}
\end{equation}
Finally the equations of motion obtained are the weakly compressible athermal Navier--Stokes equations
\begin{align}
&\p_t\rho+\uN\cdot(\rho\uu)=0,\\
&\p_t (\rho\uu)+\uN\cdot(\rho\uu\uu)=-\uN p+\uN\cdot\left(2\mu\uS\right).
\end{align}

\section{Hierarchy of non-equilibrium moments and regularization scheme}\label{sec_rec}

In this section a novel theoretical approach is proposed for the
athermal Boltzmann-BGK equation. A recursive formulation for
non-equilibrium moments is shown to exist and 
a regularization technique is proposed for the 
discrete lattice Boltzmann method.

\subsection{Recursive properties of high order moments}

The particular structure of the moments of the equilibrium distribution in absence of temperature 
allows for an elegant formulation of the high order (higher than two) non-equilibrium moments.

The Hermite coefficients of order $n$ of the equilibrium distribution can be recursively expressed as
\begin{equation}
 \ua_0^{(n)}=\ua_0^{(n-1)}\uu\hbox{, and } a_0^{(0)}=\rho.
\end{equation}
Using this relation and Eqs.~\eqref{eq_rho_euler}-\eqref{eq_j_euler} one can show that for $n\ge 3$ (see \ref{app_ce_rec} for the proof)
\begin{equation}
 a^{(n)}_{1,\al_1...\al_n}=a^{(n-1)}_{1,\al_1...\al_{n-1}}u_{\al_n}+\left(u_{\al_1}...u_{\al_{n-2}}a^{(2)}_{1,\al_{n-1}\al_n}+\hbox{perm}(\al_n)\right),\label{eq_rel_a1}
\end{equation}
where ``perm$(\al_n)$'' stands for all the cyclic index permutations of indexes from $\al_1$ to $\al_{n-1}$ ($\al_n$ is never permuted). 
One therefore notices that a Hermite coefficient of order $n$ can be expressed in terms of the the velocity and the Hermite coefficients of order order two 
and $n-1$. This property allows to reconstruct the populations up to any order by only knowing its second order coefficient
and the macroscopic velocity. 

\subsection{Discretization of the microscopic velocity space}

We notice that only the Hermite coefficients of the distribution function 
are used in the Chapman--Enskog expansion. Therefore in order to asymptotically 
recover the Navier--Stokes equations there is no need to use the complete Maxwell--Boltzmann equilibrium 
distribution but only a polynomial expansion of it. In order to discretize 
the velocity space one will use a Gauss--Hermite quadrature. The aim of this discretization
is to exactly evaluate the integral of polynomials of order $m$ with Gaussian weight as a sum
\begin{equation}
 \int\dd\uxi w(\uxi)p_m(\uxi)=\sum_{i=0}^{q-1} w_ip_m(\uxi_i),
\end{equation}
where $\{w_i\}_{i=0}^{q-1}$ and $\{\uxi_i\}_{i=0}^{q-1}$ are two sets of $q$ constant weights and
abscissae respectively.

In order to obtain asymptotically the weakly compressible limit of the BGK equation only polynomials of order $m=5$ 
need to be integrated exactly. The associated most common quadratures (see~\citet{bi_shan06}) for this case are given
by the D2Q9 (in 2D) and the D3Q15, D3Q19, and D3Q27 lattices\footnote{The D$d$Q$q$ notation denotes a lattice of dimension $d$ and with $q$ 
quadrature points} (in 3D). These quadratures allow the definition of a set of velocity discretized
distribution function noted as $\{f_i\}_{i=0}^{q-1}\equiv \{f(\ux,\uxi_i,t)\}_{i=0}^{q-1}$.
In other terms, on each position $\ux$ at time $t$ one defines $q$ independent values $\{f_i\}_{i=0}^{q-1}$.
These quantities can therefore be represented on a $q$-dimensional basis (see work of \citet{bi_dhumieres92,bi_dellar03} among others).
We emphasize here is that there are two different spaces that must be distinguished: the velocity-discretized
$q$-dimensional space and the $d$-dimensional physical space. 

The above quadratures only allow the exact representation of the populations up to second order in Hermite polynomials,
one usually truncates Eq.~\eqref{eq_hermite_eq} to order two. This means that the equilibrium
distribution is represented on a six-dimensional basis in 2D and respectively on a 10-dimensional basis in 3D, while
it is living in a 9 (for the D2Q9 lattice) or 15, 19, or 27 (for the D3Q15, D3Q19, or D3Q27 lattices) dimensional space (depending on the quadrature used). 

The BGK equation discretized in microscopic velocity space then reads
\begin{equation}
\p_t f_i+(\uxi_i\cdot\uN)f_i=-\frac{1}{\tau}\left(f_i-\fzero_i(\rho,\uu)\right).\label{eq_bgk_discr} 
\end{equation}
As pointed out above, the set populations ($f_i$ and $\fzero_i$) live in a $q$-dimensional space which means that they can be represented on a $q$-dimensional basis.
In \citet{bi_dhumieres92,bi_dhumieres02} or \citet{bi_dellar03} different bases are proposed for the expansion of the complete populations
but never used to express the equilibrium distribution. Here we propose to expand the equilibrium population on a complete basis as well. To this aim
we will use an interesting property of the D2Q9 and the D3Q27 lattices, which is that the complete 9- and 27-dimensional bases can be expressed in Hermite polynomials,
a property that does not hold for the D3Q15 and D3Q19 quadratures.
This property makes them particularly appealing in order to reuse all the calculations performed in the previous section. The distribution function
is therefore written as (in 2D)
\begin{align}
 f_i=&w_i\left(\rho+\frac{\uxi_i\cdot(\rho\uu)}{c_s^2}+\frac{1}{2c_s^4}\H^{(2)}_i:\ua^{(2)}\right.\nonumber\\
	  &\quad\quad\left.+\frac{1}{2c_s^6}\left(\H^{(3)}_{ixxy}a^{(3)}_{xxy}+\H^{(3)}_{ixyy}a^{(3)}_{xyy}\right)+\frac{1}{4c_s^8}\H^{(4)}_{ixxyy}a^{(4)}_{xxyy}\right).
\end{align}
Respectively the equilibrium and off-equilibrium parts are expanded as
\begin{align}
 \fzero_i=&w_i\rho\left(1+\frac{\uxi_i\cdot\uu}{c_s^2}+\frac{1}{2c_s^4}\H^{(2)}_i:\uu\uu\right.\nonumber\\
	  &\quad\quad\left.+\frac{1}{2c_s^6}\left(\H^{(3)}_{ixxy}u_x^2u_y+\H^{(3)}_{ixyy}u_xu_y^2\right)+\frac{1}{4c_s^8}\H^{(4)}_{ixxyy}u_x^2u_y^2\right)\label{eq_fzero_d2q9}\\
 \fone_i=&w_i\left(\frac{1}{2c_s^4}\H^{(2)}_i:\ua_1^{(2)}+\frac{1}{2c_s^6}\left(\H^{(3)}_{ixxy}a^{(3)}_{1,xxy}+\H^{(3)}_{ixyy}a^{(3)}_{1,xyy}\right)\right.\nonumber\\
	  &\quad\quad\left.+\frac{1}{4c_s^8}\H^{(4)}_{ixxyy}a^{(4)}_{1,xxyy}\right).\label{eq_fone_d2q9}
\end{align}
In 3D the equivalent expressions are given by
\begin{align}
  f_i=&w_i\left(\rho+\frac{\uxi_i\cdot(\rho\uu)}{c_s^2}+\frac{1}{2c_s^4}\H^{(2)}_i:\ua^{(2)}\right.\nonumber\\
      &\quad\quad+\frac{1}{2c_s^6}\left(\H^{(3)}_{ixxy}a^{(3)}_{xxy}+\H^{(3)}_{ixxz}a^{(3)}_{xxz}+\H^{(3)}_{ixyy}a^{(3)}_{xyy}+\H^{(3)}_{ixzz}a^{(3)}_{xzz}\right.\nonumber\\
      &\quad\quad\quad\quad\quad\quad+\left.\H^{(3)}_{iyzz}a^{(3)}_{yzz}+\H^{(3)}_{iyyz}a^{(3)}_{yyz}+2\H^{(3)}_{ixyz}a^{(3)}_{xyz}\right)\nonumber\\
      &\quad\quad+\frac{1}{4c_s^8}\left(\H^{(4)}_{ixxyy}a^{(4)}_{xxyy}+\H^{(4)}_{ixxzz}a^{(4)}_{xxzz}+\H^{(4)}_{iyyzz}a^{(4)}_{yyzz}\right.\nonumber\\
      &\quad\quad\quad\quad\quad\quad\left.+2\left(\H^{(4)}_{ixyzz}a^{(4)}_{xyzz}+\H^{(4)}_{ixyyz}a^{(4)}_{xyyz}+\H^{(4)}_{ixxyz}a^{(4)}_{xxyz}\right)\right)\nonumber\\
      &\quad\quad+\frac{1}{4c_s^{10}}\left(\H^{(5)}_{ixxyzz}a^{(5)}_{xxyzz}+\H^{(5)}_{ixxyyz}a^{(5)}_{xxyyz}+\H^{(5)}_{ixyyzz}a^{(5)}_{xyyzz}\right)\nonumber\\
      &\quad\quad\left.+\frac{1}{8c_s^{12}}\H^{(6)}_{ixxyyzz}a^{(6)}_{xxyyzz}\right),
\end{align}
\begin{align}
 \fzero_i=&w_i\rho\left(1+\frac{\uxi_i\cdot\uu}{c_s^2}+\frac{1}{2c_s^4}\H^{(2)}_i:\uu\uu\right.\nonumber\\
	  &\quad\quad+\frac{1}{2c_s^6}\left(\H^{(3)}_{ixxy}u_x^2u_y+\H^{(3)}_{ixxz}u_x^2u_z+\H^{(3)}_{ixyy}u_xu_y^2+\H^{(3)}_{ixzz}u_xu_z^2\right.\nonumber\\
	  &\quad\quad\quad\quad\quad\quad\left.+\H^{(3)}_{iyzz}u_yu_z^2+\H^{(3)}_{iyyz}u_y^2u_z+2\H^{(3)}_{ixyz}u_xu_yu_z\right)\nonumber\\
	  &\quad\quad+\frac{1}{4c_s^8}\left(\H^{(4)}_{ixxyy}u_x^2u_y^2+\H^{(4)}_{ixxzz}u_x^2u_z^2+\H^{(4)}_{iyyzz}u_y^2u_z^2\right.\nonumber\\
	  &\quad\quad\quad\quad\quad\quad\left.+2\left(\H^{(4)}_{ixyzz}u_xu_yu_z^2+\H^{(4)}_{ixyyz}u_xu_y^2u_z+\H^{(4)}_{ixxyz}u_x^2u_yu_z\right)\right)\nonumber\\
	  &\quad\quad+\frac{1}{4c_s^{10}}\left(\H^{(5)}_{ixxyzz}u_x^2u_yu_z^2+\H^{(5)}_{ixxyyz}u_x^2u_y^2u_z+\H^{(5)}_{ixyyzz}u_xu_y^2u_z^2\right)\nonumber\\
	  &\quad\quad\left.+\frac{1}{8c_s^{12}}\H^{(6)}_{ixxyyzz}u_x^2u_y^2u_z^2\right),\label{eq_fzero_d3q27}
\end{align}
\begin{align}
  \fone_i=&w_i\left(\frac{1}{2c_s^4}\H^{(2)}_i:\ua_1^{(2)}\right.\nonumber\\
	  &\quad\quad+\frac{1}{2c_s^6}\left(\H^{(3)}_{ixxy}a^{(3)}_{1,xxy}+\H^{(3)}_{ixxz}a^{(3)}_{1,xxz}+\H^{(3)}_{ixyy}a^{(3)}_{1,xyy}+\H^{(3)}_{ixzz}a^{(3)}_{1,xzz}\right.\nonumber\\
	  &\quad\quad\quad\quad\quad\quad\left.+\H^{(3)}_{iyzz}a^{(3)}_{1,yzz}+\H^{(3)}_{iyyz}a^{(3)}_{1,yyz}+2\H^{(3)}_{ixyz}a^{(3)}_{1,xyz}\right)\nonumber\\
	  &\quad\quad+\frac{1}{4c_s^8}\left(\H^{(4)}_{ixxyy}a^{(4)}_{1,xxyy}+\H^{(4)}_{ixxzz}a^{(4)}_{1,xxzz}+\H^{(4)}_{iyyzz}a^{(4)}_{1,yyzz}\right.\nonumber\\
	  &\quad\quad\quad\quad\quad\quad\left.+2\left(\H^{(4)}_{ixyzz}a^{(4)}_{1,xyzz}+\H^{(4)}_{ixyyz}a^{(4)}_{1,xyyz}+\H^{(4)}_{ixxyz}a^{(4)}_{1,xxyz}\right)\right)\nonumber\\
	  &\quad\quad+\frac{1}{4c_s^{10}}\left(\H^{(5)}_{ixxyzz}a^{(5)}_{1,xxyzz}+\H^{(5)}_{ixxyyz}a^{(5)}_{1,xxyyz}+\H^{(5)}_{ixyyzz}a^{(5)}_{1,xyyzz}\right)\nonumber\\
	  &\quad\quad\left.+\frac{1}{8c_s^{12}}\H^{(6)}_{ixxyyzz}a^{(6)}_{1,xxyyzz}\right).\label{eq_fone_d3q27}
\end{align}
The Hermite coefficients of the equilibrium distribution are the ones obtained 
from of the continuous equilibrium distribution for both the 2D and 3D cases, simplifying
the computations. Another important remark is that not all the Hermite polynomials are used
at each order. This is due to the fact that the quadrature is not accurate enough to represent 
exactly all the Hermite polynomials, but only the ones that are used in the formulas above.

\subsection{Chapman--Enskog expansion of the model}\label{sec_ce_mod}

The Chapman--Enskog expansion of this model asymptotically leads to the following constitutive equation for $\ua^{(2)}_1$
in three dimensions
\begin{align}
 a_{1,xx}^{(2)} &= -2c_s^2\rho\tau S_{xx}+\underbrace{\tau\p_x(\rho u_x^3)}_{\ast},\label{eq_axx_sxx_complete}\\
 a_{1,xy}^{(2)} &= -2c_s^2\rho\tau S_{xy},\\
 a_{1,xz}^{(2)} &= -2c_s^2\rho\tau S_{xz},\\
 a_{1,yy}^{(2)} &= -2c_s^2\rho\tau S_{yy}+\underbrace{\tau\p_y(\rho u_y^3)}_{\ast},\\
 a_{1,yz}^{(2)} &= -2c_s^2\rho\tau S_{yz},\\
 a_{1,zz}^{(2)} &= -2c_s^2\rho\tau S_{zz}+\underbrace{\tau\p_z(\rho u_z^3)}_{\ast},\label{eq_azz_szz_complete}
\end{align}
where the ``$\ast$'' terms are $\O(\Ma^3)$ order error terms which are not present
in the continuous case (see Eq.~\eqref{eq_a1_2-classical}). In the case of the order two standard BGK model (where $\fzero_i$ is only expanded up to order two in Hermite polynomials) 
it reads
\begin{align}
 a_{1,xx}^{(2)} &= -2c_s^2\rho\tau S_{xx}+\tau(\p_x(\rho u_x^3)\underbrace{+\p_y(\rho u_yu_x^2)+\p_z(\rho u_zu_x^2)}_{\ast\ast}),\label{eq_axx_sxx}\\
 a_{1,xy}^{(2)} &= -2c_s^2\rho\tau S_{xy}\underbrace{+\tau(\p_x(\rho u_x^2 u_y)+\p_y(\rho u_x u_y^2)+\p_z(\rho u_x u_y u_z))}_{\ast\ast},\\
 a_{1,xz}^{(2)} &= -2c_s^2\rho\tau S_{xz}\underbrace{+\tau(\p_y(\rho u_x u_y u_z)+\p_x(\rho u_z u_x^2)+\p_z(\rho u_x u_z^2))}_{\ast\ast},\\
 a_{1,yy}^{(2)} &= -2c_s^2\rho\tau S_{yy}+\tau(\p_y(\rho u_y^3)\underbrace{+\p_x(\rho u_xu_y^2)+\p_z(\rho u_zu_y^2)}_{\ast\ast}),\\
 a_{1,yz}^{(2)} &= -2c_s^2\rho\tau S_{yz}\underbrace{+\tau(\p_x(\rho u_x u_y u_z)+\p_y(\rho u_z u_y^2)+\p_z(\rho u_y u_z^2))}_{\ast\ast},\\
 a_{1,zz}^{(2)} &= -2c_s^2\rho\tau S_{zz}+\tau(\p_z(\rho u_z^3)\underbrace{+\p_x(\rho u_xu_z^2)+\p_y(\rho u_yu_z^2)}_{\ast\ast}),\label{eq_azz_szz}
\end{align}
where the ``$\ast\ast$'' terms are the terms that are not present anymore in the new model. One can see that in our case
the non-diagonal terms are exact.

Furthermore since we now expand the distribution function up to a limited order in Hermite polynomials
the relations of Eq.~\eqref{eq_rel_a1} are not anymore exactly verified. 
Nevertheless the difference between the exact relation and the error committed is compatible with the low compressibility approximation 
of the scheme (low Mach number approximation). In other terms Eq.~\eqref{eq_rel_a1} reads in the discrete case 
\begin{align}
 a^{(n)}_{1,\al_1...\al_n}&=a^{(n-1)}_{1,\al_1...\al_{n-1}}u_{\al_n}+\left(u_{\al_1}...u_{\al_{n-2}}a^{(2)}_{1,\al_{n-1}\al_n}+\hbox{perm}(\al_n)\right)\nonumber\\
 &\quad\quad+\O(\Ma^{n+1}),\label{eq_rel_a1_lowma}
\end{align}
where $\O$ stands for the order of the error committed. These terms are spurious 
terms that are due to quadrature (discretization in velocity space) errors in the expansion
and are assumed to be small because they are one order higher in physical velocity.
For a more detailed expression for the $\O(\Ma^{n+1})$ terms see \ref{app_hermite}.

\subsection{Time-space discretization}\label{subsec_time_space}

The time space discretization of Eq.~\eqref{eq_bgk_discr} is done as usual by integrating it along characteristics with 
the trapezoidal rule and making the following change of variables (see \citet{bi_dellar01} for example)
\begin{equation}
 \fb_i=\fb_i+\frac{1}{2\tau}\left(\fb_i-\fzero_i\right).
\end{equation}
This leads to the following lattice Boltzmann method scheme
\begin{equation}
 \fb_i(\ux+\uxi_i,t+1)=\fb_i(\ux,t)-\frac{1}{\bar{\tau}}\left(\fb_i(\ux,t)-\fzero_i(\ux,t)\right),\label{eq_full_discr_bgk}
\end{equation}
where $\bar{\tau}\equiv \tau+1/2$. From now on the ``bar'' is omitted for brevity in the notations.

\subsection{Regularization scheme}

The numerical scheme used here is given by Eq.~\eqref{eq_full_discr_bgk} where $f_i$ is ``regularized'' at each iteration 
as done in \citet{bi_latt06}
\begin{equation}
 f_i=\fzero_i+\fone_i,\label{eq_reg}
\end{equation}
where $\fone_i$ is computed with Eqs.~\eqref{eq_fone_d2q9} or~\eqref{eq_fone_d3q27} depending on the dimension of the physical space. 
Furthermore the Hermite coefficients of Eqs.~\eqref{eq_fone_d2q9} and~\eqref{eq_fone_d3q27} ($\ua_1^{(n)}$, with $n>2$) are evaluated thanks to the recursive formulation of Eq.~\eqref{eq_rel_a1}.
For efficiency reasons Eq.~\eqref{eq_full_discr_bgk} can thus be rewritten as
\begin{equation}
 f_i(\ux+\uxi_i,t+1)=\fzero_i+\left(1-\frac{1}{\tau}\right)\fone_i.\label{eq_discr_bgk_reg}
\end{equation}
This is the novel scheme proposed in this paper. The model will be validated in Sec.~\ref{sec_bench}
and a comparison with an existing multiple-relaxation-time model will be performed.

\subsection{Von Neumann linear stability analysis}\label{subsec_neumann}

The aim of this section is to perform the linear stability analysis of the scheme. 
More details about the Von Neumann stability analysis of the lattice Boltzmann method can be found among others
in \citet{bi_lallemand00,bi_ricot09,bi_xu11,bi_xu12}.

By decomposing the distribution function into the sum of a stationary part (noted $\bar{f}_i$ which must not be confused with the $\bar{f}$ of
Subsec.~\ref{subsec_time_space}) and a small fluctuating part, noted $f'_i$,
\begin{equation}
 f_i=\bar{f}_i+f'_i,
\end{equation}
and by defining $\Omega_i$ as the r.h.s. of Eq.~\eqref{eq_discr_bgk_reg} 
\begin{equation}
\Omega_i\equiv \fzero_i+\left(1-\frac{1}{\tau}\right)\fone_i,
\end{equation}
the linearized lattice Boltzmann scheme is found to be given by
\begin{equation}
 f_i'(\ux+\uxi_i,t+1)=\sum_j\Lambda_{ij}f_j'(\ux,t),\label{eq_lin1}
\end{equation}
where $\uLambda$ is defined as
\begin{equation}
 \Lambda_{ij}=\left.\frac{\p \Omega_i}{\p f_j}\right|_{f_j=\bar{f}_j}.
\end{equation}
Working in Fourier space and looking for plane wave solutions Eq.~\eqref{eq_lin1}
becomes
\begin{equation}
 f_l'(\uk,t+1)=\sum_{j,k}A^{-1}_{lj}\Lambda_{jk}f_k'(\uk,t),\label{eq_lin2}
\end{equation}
where $A^{-1}_{jk}=\delta_{jk}\exp{(-i \uk\cdot\uxi_j)}$, with $\delta_{jk}$ the Kronecker symbol and $i=\sqrt{-1}$.

The eigenvalues, $\lambda_j$, of the matrix $\uA^{-1}\uLambda$ allow to obtain the dispersion-dissipation
relations of the numerical scheme, $\omega_j(\uk)=i\log{\lambda_j}$. While an analytic approach is used to determine 
these eigenvalues in \citet{bi_lallemand00}, here we simply used a linear algebra package to determine
numerically these eigenvalues.

The Von Neumann stability analysis of the Navier--Stokes gives the following dispersion-dissipation relations
\begin{align}
 \Re(\omega_\pm)&=\pm||\uk||c_s+\uk\cdot\uu,\\
 \Im(\omega_\pm)&=-||\uk||^2\frac{1}{2}\left(\frac{2D-2}{D}\nu+\eta\right),\\
 \Re(\omega_s)&=\uk\cdot\uu,\\
 \quad \Im(\omega_s)&=-||\uk||^2\nu,
\end{align}
where $\eta=\frac{2}{D}\nu$ for the BGK model.

\begin{figure}
\begin{center}
\includegraphics[width=.45\textwidth]{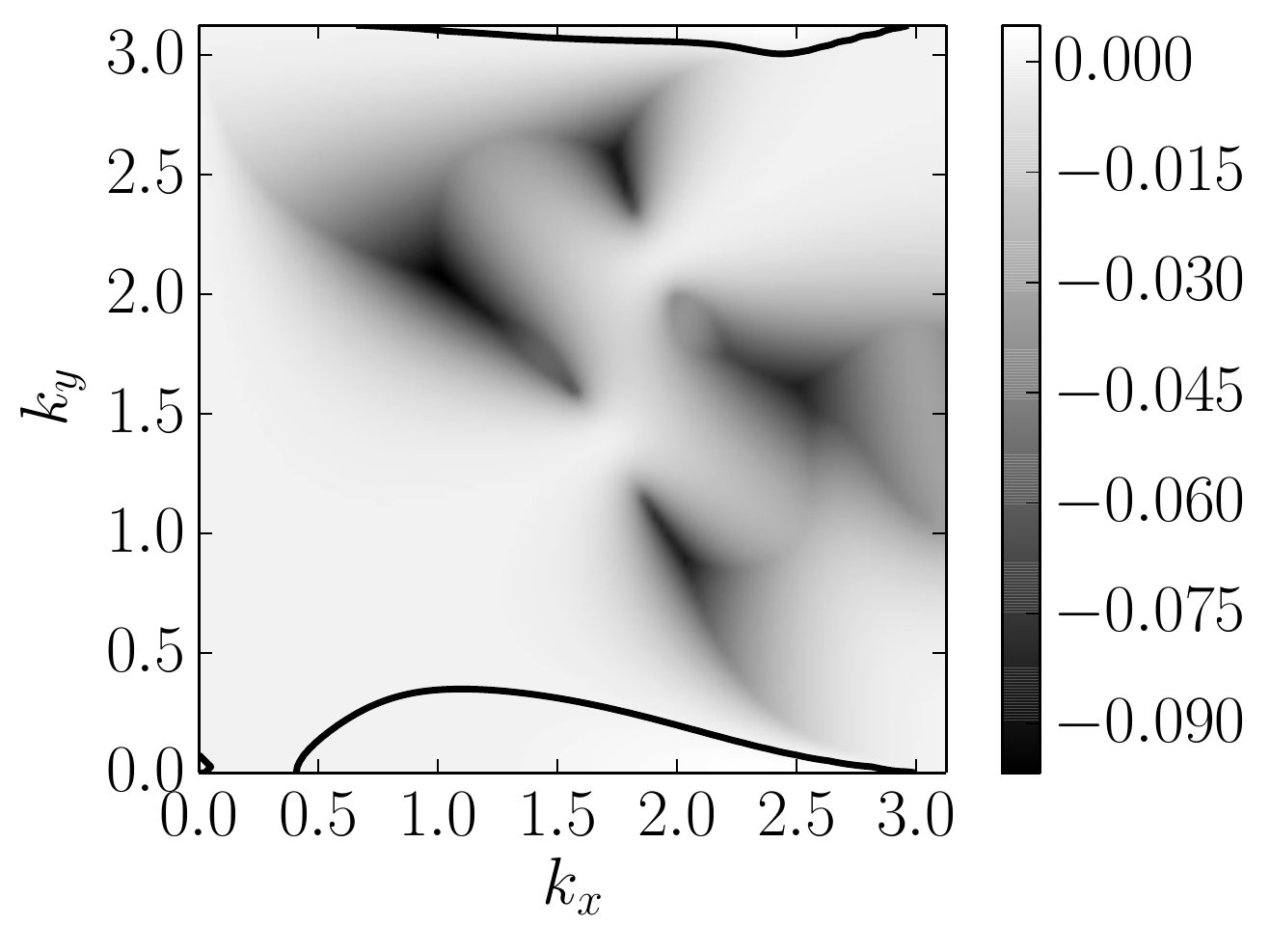}
\includegraphics[width=.45\textwidth]{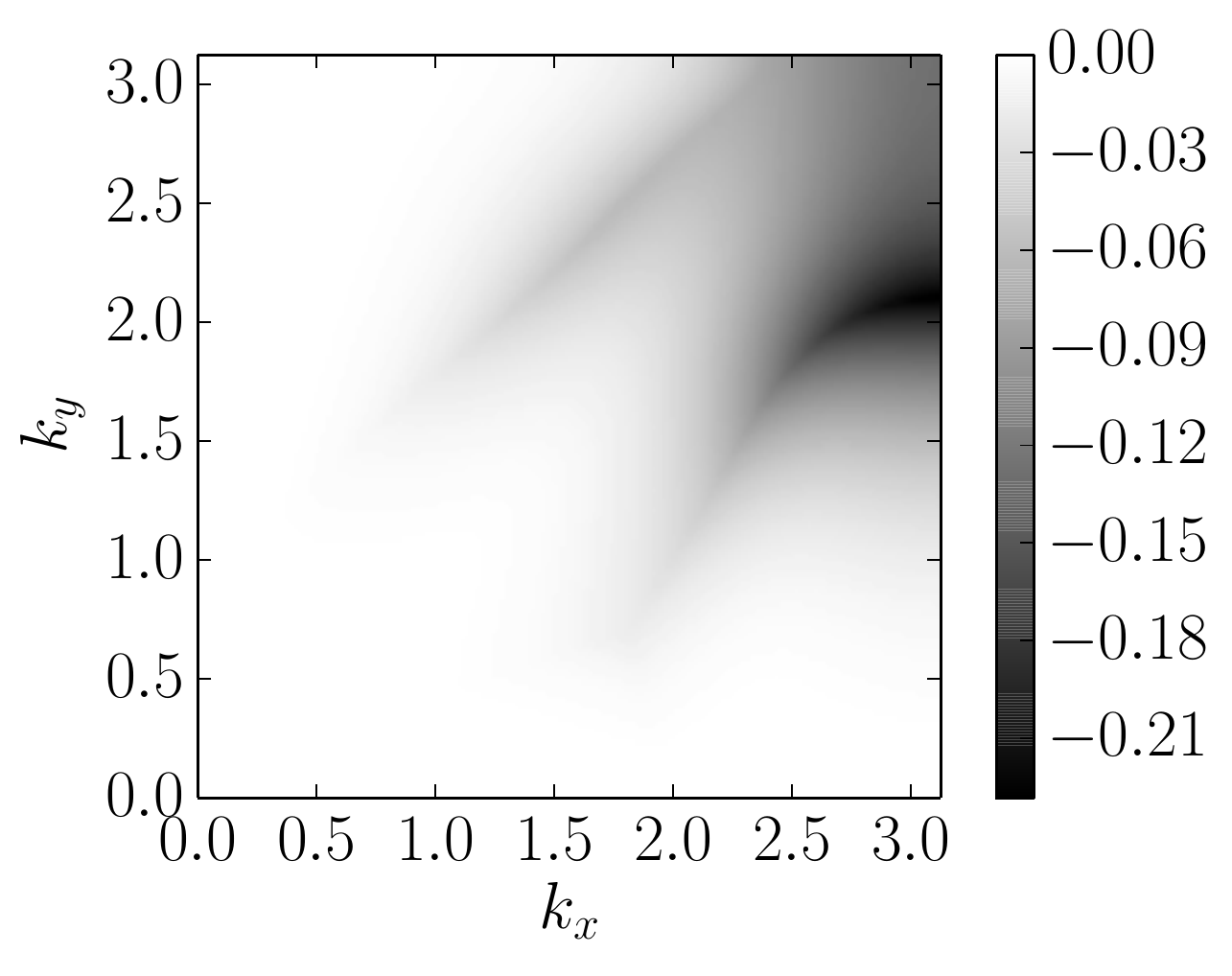}
\caption{Maximal value of the dissipation, $\max_j \omega_j(\uk)$, for the MRT (left) and present (right) models with  $u_x=0.2$, 
for the D2Q9 lattice, and $\tau=0.5001$. The solid line represents the isocontour where $\max_j \omega_j(\uk)=0$. }\label{fig_disp_diss_map}
\end{center}
\end{figure}

\begin{figure}
\begin{center}
\includegraphics[width=.45\textwidth]{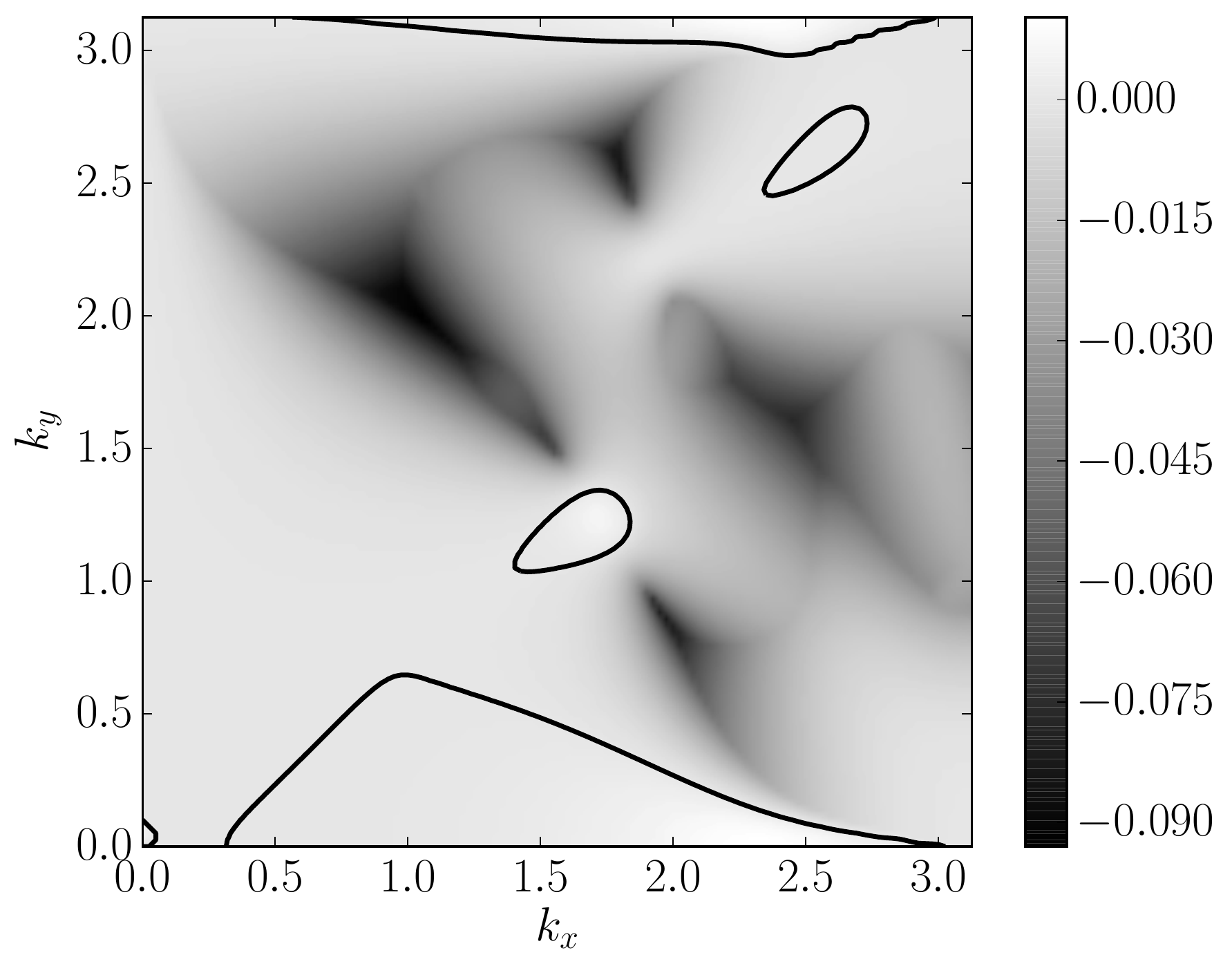}
\includegraphics[width=.45\textwidth]{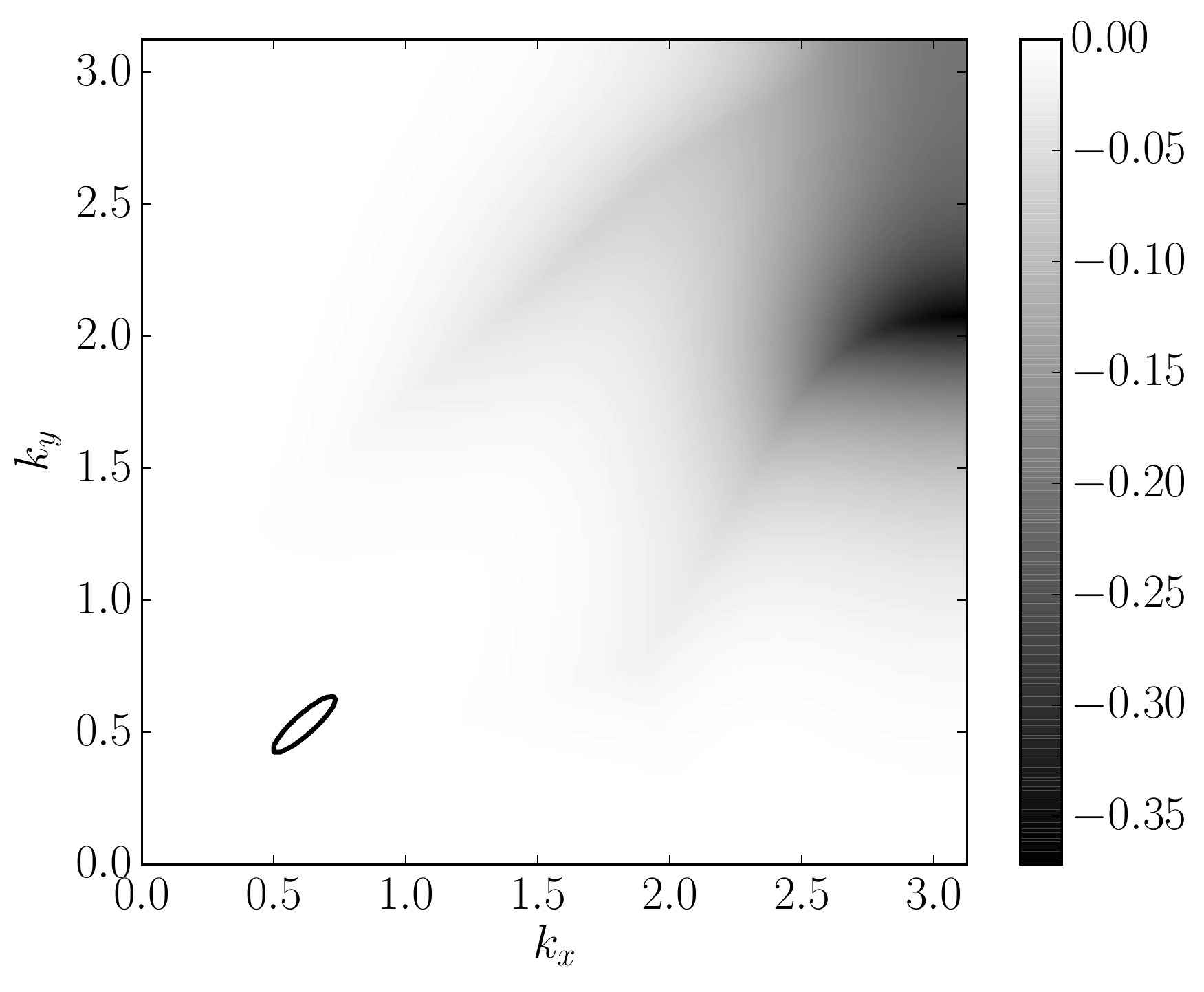}
\caption{Maximal value of the dissipation, $\max_j \omega_j(\uk)$, for the MRT (left) and present (right) models with  $u_x=0.248$, 
for the D2Q9 lattice, and $\tau=0.5001$. The solid line represents the isocontour where $\max_j \omega_j(\uk)=0$. }\label{fig_disp_diss_map_ma}
\end{center}
\end{figure}

We now compare the eigenvalues of the collision operator of the present model
we the ones obtained for the MRT model proposed by \citet{bi_lallemand00}. 
The stability analysis depicted in this section has been performed for $\tau=0.5001$ and $\uu=(0.2,0)$ (corresponding to $\Ma=0.346$). 
\begin{figure}
\begin{center}
\includegraphics[width=.75\textwidth]{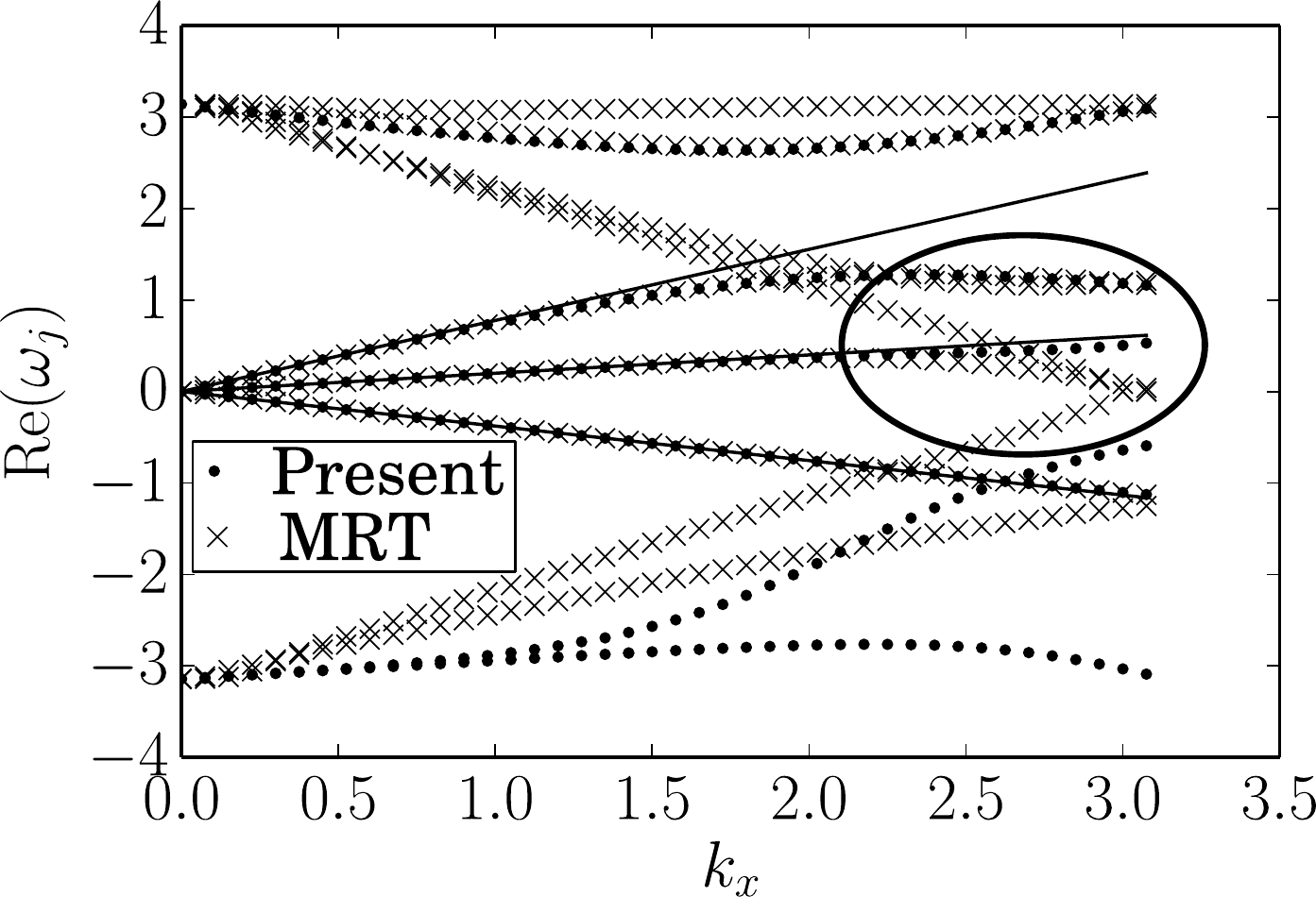}
\caption{Dispersion with respect to $k_x$ ($k_y=0$) for the present model and the MRT model with the $u_x=0.2$, 
for the D2Q9 lattice, and $\tau=0.5001$. This circled region highlights the region where 
there is a significant improvement of the dispersion relation of the present model as compared to the MRT model.}\label{fig_disp_kx}
\end{center}
\end{figure}
\begin{figure}
\begin{center}
\includegraphics[width=.75\textwidth]{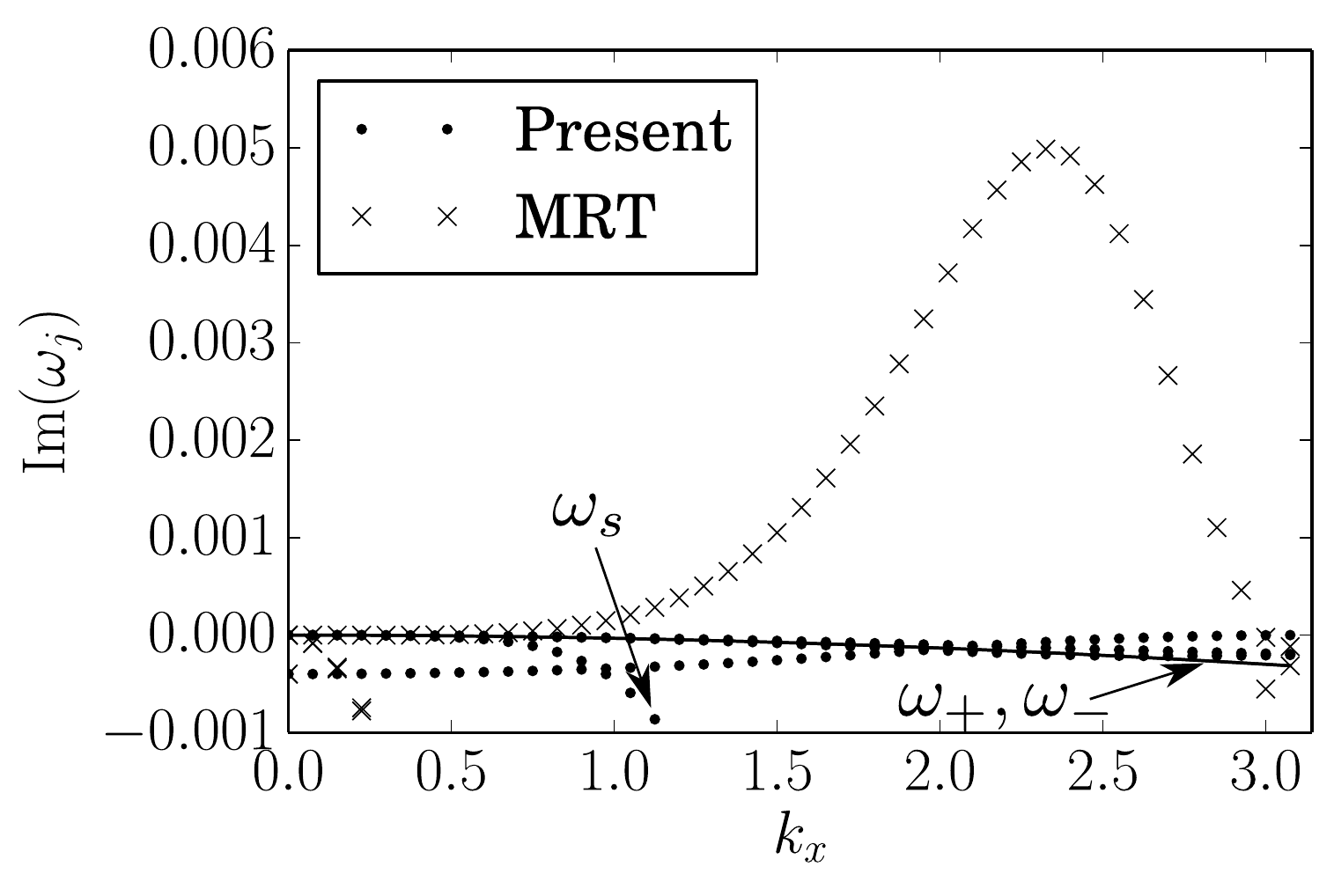}
\caption{Dissipation (right) with respect to $k_x$ ($k_y=0$) for the present model and the MRT model with the $u_x=0.2$, 
for the D2Q9 lattice, and $\tau=0.5001$.}\label{fig_diss_kx}
\end{center}
\end{figure}
As one can see from Fig.~\ref{fig_disp_kx} the dispersion relations for the present scheme are relatively similar to the 
ones obtained with an MRT approach except for the shear mode where the dispersion relation remains 
very close to the Navier--Stokes result until $k_x=\pi$ which is not the case for the MRT (see the circled region 
of Fig.~\ref{fig_disp_kx}). For the dissipation (see Fig.~\ref{fig_diss_kx})
one clearly sees that for $k_y=0$ there is no unstable mode for the present model ($\Im(\omega_j)<0,\forall k_x,j$) while 
this is not the case for the MRT model. Furthermore one can also notice that while the acoustic modes are only weakly dissipated,
except for $\omega_s$,
(and therefore the proposed scheme could be very suitable for aeroacoustic simulations) the other spurious modes
are dissipated very fast and therefore the scheme should suffer of less numerical instabilities. 
The increased linear stability of the model is depicted in Fig.~\ref{fig_disp_diss_map} where one can see that 
the imaginary part of the eigenvalues of the numerical scheme is always negative and therefore the scheme 
has an unconditional linear stability. This feature is not present in the case of the MRT model as for some 
values of $\uk$ the imaginary part of the eigenvalues of the evolution operator become positive. 
Of course
as one increases the magnitude of the velocity unstable modes will start to appear. The unstable modes start to appear at $u_x=0.248$ 
(corresponding to $\Ma=0.43$)
for $\tau=0.5001$  as shown on Fig.~\ref{fig_disp_diss_map_ma}.

\subsection{Boundary conditions}

The aim of this subsection is not to give a detailed view of the way to implement boundary conditions
since this topic is extensively treated in the literature (see \citet{bi_latt08,bi_malaspinas11,bi_zou-he97,bi_inamuro95,bi_skordos93} among others for some references).
It rather explains how the proposed regularization model is compatible with all the cited boundary conditions.

One way to deal with Dirichlet boundary conditions in the lattice Boltzmann method is to use the regularized procedure described in \citet{bi_latt08} for example. 
The generic idea is to impose a velocity $\uu_\mathrm{bc}$ at the boundary. First one uses the symmetries of the lattice
to compute $\rho$. Then it is possible to compute $\fzero_i$ (which only depends on $\rho$ and $\uu_\mathrm{bc}$). Finally 
$\uP^{(1)}$ can be computed using a finite difference scheme through $\uS$ or by using the symmetries of the lattice (see~\cite{bi_latt08}). 
Then $\fone_i$ is computed by using the following formula
\begin{equation*}
\fone_i=\frac{w_i}{2c_s^4}\H^{(2)}_i:\uP^{(1)}.
\end{equation*}
We notice that this formula is actually exact if $\uu_\mathrm{bc}=0$. By replacing 
$\uu_\mathrm{bc}$ by zero in Eq.~\eqref{eq_fone_d2q9} or~\eqref{eq_fone_d3q27} one is simply left with
the equation above. Then if $\uu_\mathrm{bc}\neq 0$, the procedure is exactly the same as for the two strategies
discussed above and we should simply use Eqs.~\eqref{eq_fone_d2q9} or~\eqref{eq_fone_d3q27} for the computation of $\fone_i$.
Finally one simply replaces all the populations on the boundary mesh point with the regularization formula~\eqref{eq_reg}.

\subsection{Differences with existing stabilization techniques}

Apart from the MRT-LBM there exists different techniques to increase the stability and accuracy 
of the BGK-LBM scheme. In this subsection we discuss the major distinctions between the present scheme 
and some of these approaches. We will limit the discussion to the regularization, entropic, and selective
viscosity filter techniques.

\subsubsection{The regularized model}
\begin{figure}
\begin{center}
\includegraphics[width=.75\textwidth]{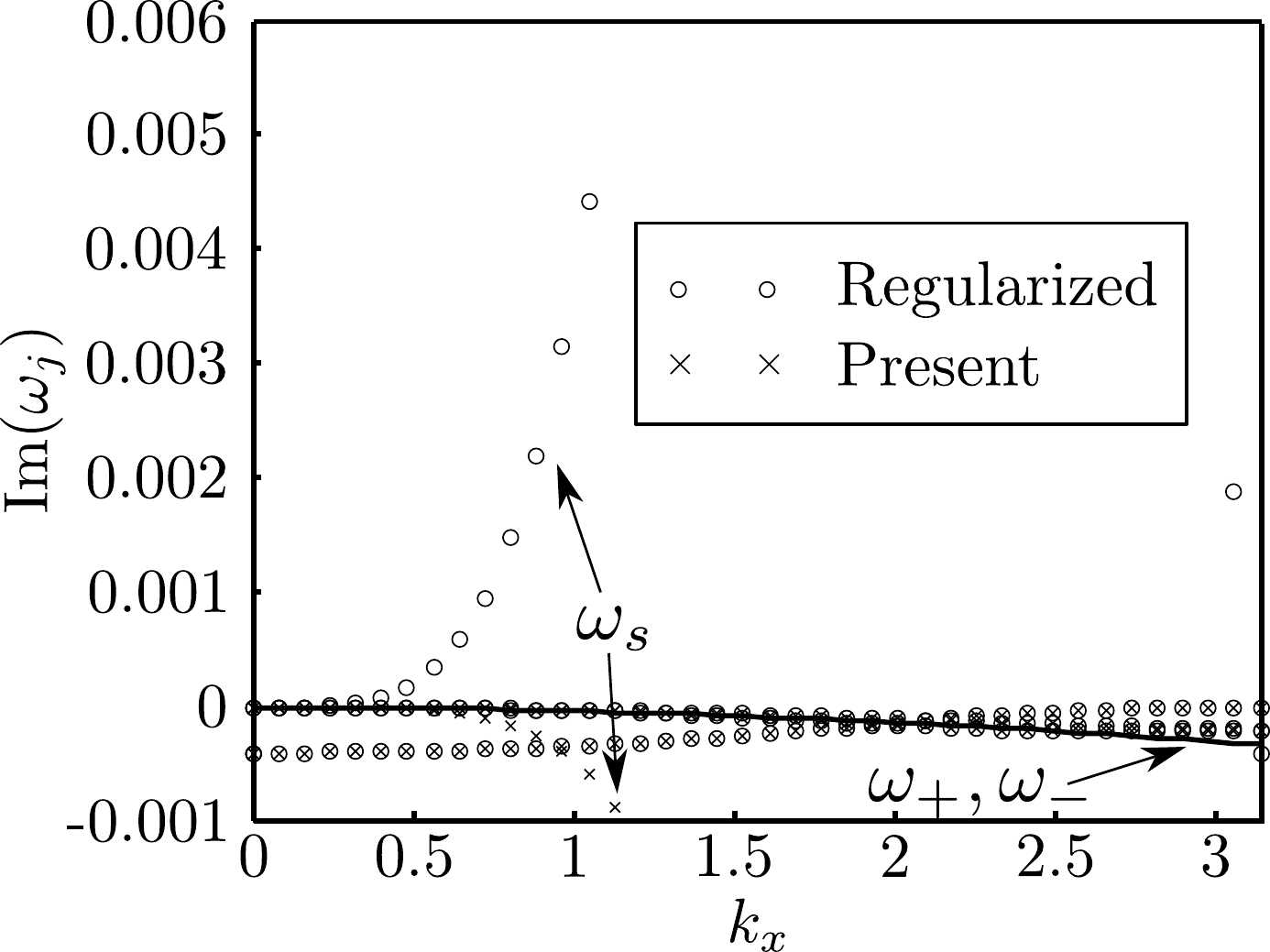}
\caption{Dissipation (right) with respect to $k_x$ ($k_y=0$) for the present model and the MRT model with the $u_x=0.2$, 
for the D2Q9 lattice, and $\tau=0.5001$.}\label{fig_diss_kx_reg}
\end{center}
\end{figure}

The existing class of regularized models (see \citet{bi_zhang06,bi_latt06}) belongs to the same family as the model
presented here. The general idea is the same as the one used for the present model. The collision operator 
is the same as Eq.~\eqref{eq_discr_bgk_reg} 
\begin{equation*}
f_i(\ux+\uxi_i,t+1)=\fzero_i+\left(1-\frac{1}{\tau}\right)\fone_i,
\end{equation*}
but instead of using the equilibrium distribution of Eqs.~\eqref{eq_fzero_d2q9} (in 2D) 
or \eqref{eq_fzero_d3q27} (in 3D), and off-equilibrium distribution of Eqs.~\eqref{eq_fone_d2q9} (in 2D) 
or \eqref{eq_fone_d3q27} (in 3D), one truncates the series at order two in Hermite polynomials, which amounts to use
\begin{align}
\fzero_i=&w_i\rho\left(1+\frac{\uxi_i\cdot\uu}{c_s^2}+\frac{1}{2c_s^4}\H^{(2)}_i:\uu\uu\right),\label{eq_feq_2}\\
\fone_i=&\frac{w_i}{2c_s^4}\H^{(2)}_i:\ua_1^{(2)}.
\end{align}
This regularization technique removes all the moments of order higher than two in Hermite polynomials 
from the distribution function. These are considered as negligible in the asymptotic limit of the weakly compressible 
Navier--Stokes equation. The removal of these higher order terms affect the accuracy 
of the constitutive equation for the stress tensor $\ua^{(2)}_1=\uP^{(1)}$ as shown in Eqs.~\eqref{eq_axx_sxx_complete}-\eqref{eq_azz_szz_complete}
and Eqs.~\eqref{eq_axx_sxx}-\eqref{eq_azz_szz}. The present regularization not only provides a
more accurate constitutive equation for the deviatoric stress but also preserves 
the recursive relation of the $\ua_1^{(n)}$ (for $n\geq 3$) terms. These differences lead
to a major difference for the linear stability analysis of the two schemes. A comparison for of the dissipation 
(see Subsec.~\ref{subsec_neumann}) for $u_x=0.2$ and $\tau=0.5001$ is shown in Fig.~\ref{fig_diss_kx_reg}.
One can see that while the $\omega_\pm$ eigenvalues have very similar values for both models,
the difference lies in the $\omega_s$ eigenvalue. 
For the regularized model this eigenvalue is positive (and therefore an 
unstable mode exists) while it is negative for the present model. Therefore one expects
the present model to exhibit a much more stable behavior.

\subsubsection{The entropic model}

The entropic model (see among other \citet{bi_ansumali02,bi_boghosian03,bi_chikatamarla06}) is based on a different 
philosophy for the construction of the numerical scheme. The major difference is the existence of  
an $H$-function defined as
\begin{equation}
 H=\sum_i f_i\log{\frac{f_i}{w_i}}.
\end{equation}
The assumption is then made that there exists a discrete $H$-theorem which states that
\begin{enumerate}
 \item The equilibrium distribution, $\fzero_i$ minimizes the $H$ function under the constrains that $\sum_i \fzero_i=\rho$ and $\sum_i \uxi_i \fzero_i=\rho\uu$.
 \item The $H$ function is monotonically decreasing.
\end{enumerate}
This second condition is imposed through the following collision operator
\begin{equation}
 f(\ux+\uxi_i,t+1)=f_i-\frac{\alpha}{2\tau}\left(f_i-\fzero_i\right),\label{eq_entropic_collision}
\end{equation}
where $\alpha>0$ is computed such that
\begin{equation}
 H(f_i-\alpha(f_i-\fzero_i))=H(f_i).
\end{equation}
The collision operator of the entropic model guarantees an unconditional stability of the scheme. This comes nevertheless at a high computational
cost since at each point and at each time the above non-linear implicit equation must be solved. As shown in several 
references (see \citet{bi_malaspinas08} among others) the presence of the $\alpha$ parameter has as an effect to locally increase the viscosity (and therefore the dissipation).
Therefore one would expect that the dissipation of the entropic scheme would be more important 
and therefore less suitable for acoustic propagation for example.

\subsubsection{The selective viscosity model}

In the selective viscosity model proposed by \citet{bi_ricot09} the basic idea is to use the
standard BGK-LBM collision operator (see Eq.~\eqref{eq_full_discr_bgk}) and to 
increase the stability of the model
by applying a low-pass filter on the $f_i$ at each point and at each time step. The filtering operation is defined as
\begin{equation}
 \langle f_i(\ux,t)\rangle =f_i(\ux,t)-\sigma\sum_{j=1}^D\sum_{n=-N}^N d_n f_i(\ux+n\ue_j),
\end{equation}
where $\ue_j$ are the $D$ unit basis vectors of the Cartesian coordinate system, the $d_n$ are
the $2N+1$ filter coefficients, and $\sigma\in[0,1]$ is the strength of the filter.
This filtering operation removes the high frequency oscillations responsible for the numerical 
instabilities appearing in the model. The different filters proposed in~\cite{bi_ricot09}
involve non-local computations (the filters have width between three and nine 
mesh points) that impacts greatly the computational efficiency of the scheme since not only more operations must be performed
at each mesh point but also the amount of communications (which are crucial for parallel efficiency) is also increased.
In the model presented here no such non-local operations are performed reducing
the computational cost with respect to the selective viscosity models.

Furthermore the filtering operation implies the existence of a cutoff which removes the high wavenumber
components of the flow. As shown in \citet{bi_ricot09} 
the large wavenumber dissipation is greatly enhanced in order to stabilize the numerical scheme. 
All the $\Im(\omega_j)$ in the Von Neumann stability analysis 
are greatly decreased after the cutoff value which decreases the accuracy of the 
propagation of high wavenumber components of the flow. 
In the present model only the dissipation of $\omega_s$ is increased and $\omega_\pm$ 
are left untouched with respect to the LBM-BGK scheme which should 
provide a better accuracy for acoustic applications.

\section{Benchmarks}\label{sec_bench}

In order to validate the model we are going to study a 2D and a 3D case, namely the dipole-wall interaction, and the turbulent jet. 
Both these flows are challenging from the numerical point of view since they exhibit a turbulent behavior (2D as well as 3D turbulence). 

\subsection{Dipole--wall interaction}

This benchmark is based on the works of \citet{bi_clercx06} and \cite{bi_latt07}. It analyzes the time evolution of a self-propelled dipole confined 
within a 2D box. The geometry of the box is a square domain $[-L,L]\times[-L,L]$,  surrounded by no-slip walls. 
The initial condition describes two counter-rotating monopoles, one with positive core vorticity at the position $(x_1,y_1)$ and 
the other one with negative core vorticity at $(x_2,y_2)$. This is obtained with an initial velocity field $\uu_0=(u_x,u_y)$ which reads as follows in dimensionless variables
\begin{align}
u_x &= -\frac{1}{2}\left\|\omega_e\right\|(y-y_1)\mathrm{e}^{-(r_1/r_0)^2}+\frac{1}{2}\left\|\omega_e\right\|(y-y_2)\mathrm{e}^{-(r_2/r_0)^2},\label{eq_dipole1}\\
u_y &= +\frac{1}{2}\left\|\omega_e\right\|(x-x_1)\mathrm{e}^{-(r_1/r_0)^2}-\frac{1}{2}\left\|\omega_e\right\|(x-x_2)\mathrm{e}^{-(r_2/r_0)^2}.\label{eq_dipole2}
\end{align}
Here, $r_i=\sqrt{(x-x_i)^2+(y-y_i)^2}$, defines the distance to the monopole centers. The parameter $r_0$ labels the diameter of a monopole and $\omega_e$ its core vorticity.

The quantity we are interested in monitoring is the average enstrophy which is defined by
\begin{equation}
\left\langle\Omega\right\rangle(t) = \frac{1}{2}\int_{-1}^{1}\int_{-1}^{1}\omega^2(\bm{x},t)\dd x\dd y,
\end{equation}
where $\omega=\partial_x u_y - \partial_y u_x$ is the flow vorticity. 
\begin{figure}
\begin{center}
\includegraphics[width=.6\textwidth]{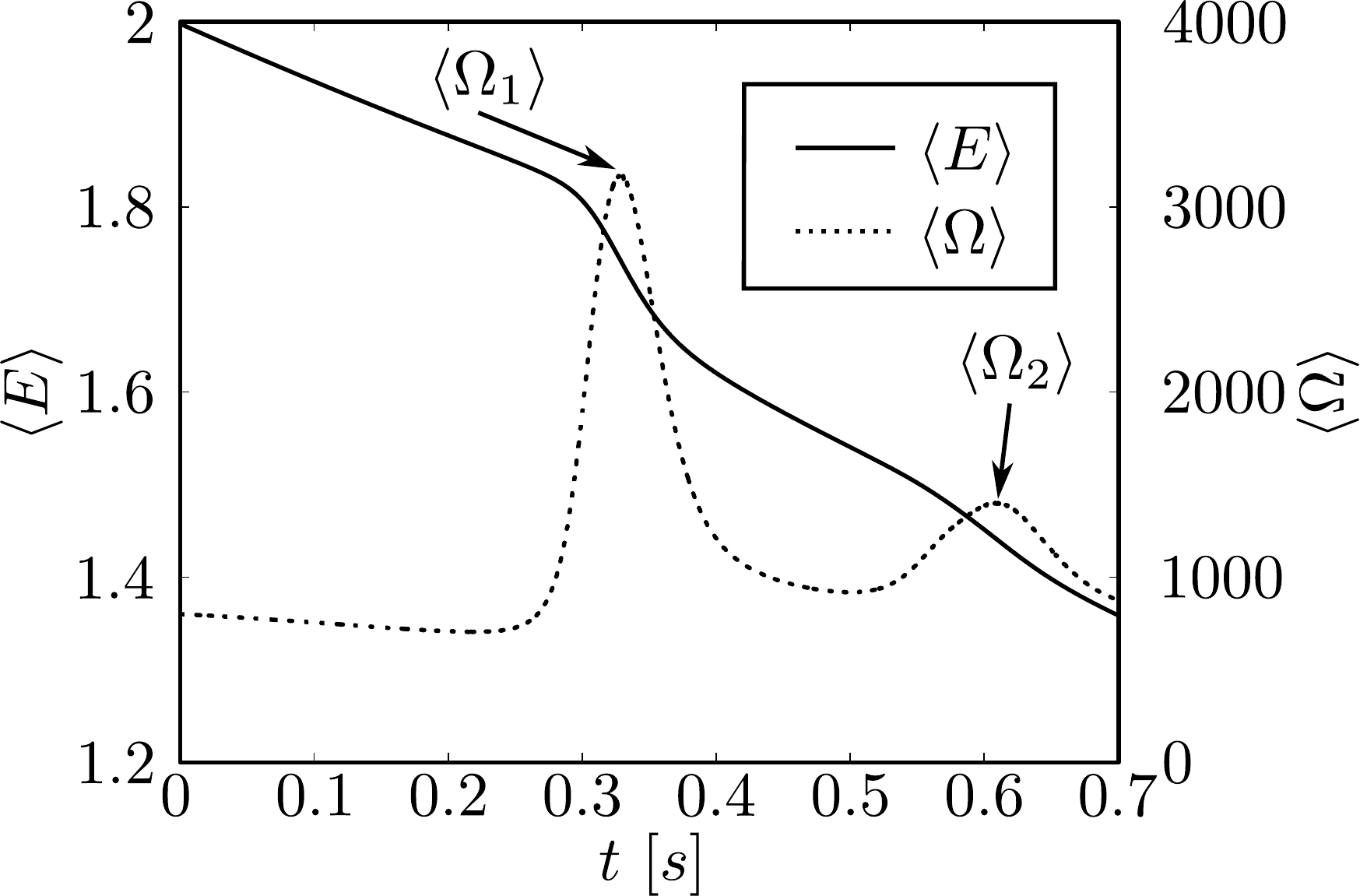}
\caption{Average energy (plain line) and average enstrophy (dotted line) evolution with time. The two enstrophy peaks 
$\langle\Omega_1\rangle$ and $\langle\Omega_2\rangle$ are highlighted.}\label{fig_energy_enstrophy}
\end{center}
\end{figure}

\begin{figure}
\begin{center}
\includegraphics[width=.24\textwidth]{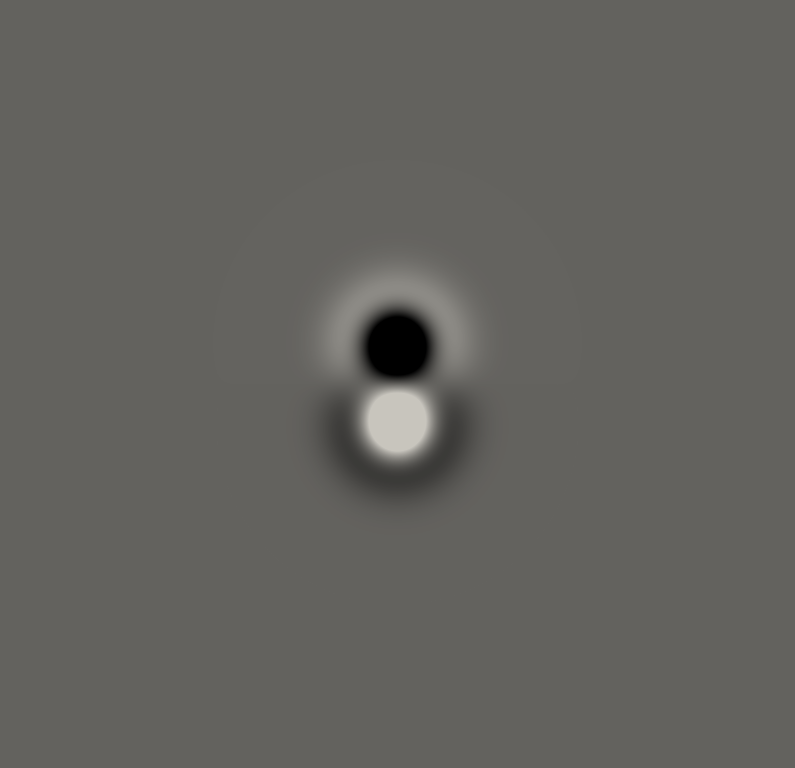}\ \includegraphics[width=.24\textwidth]{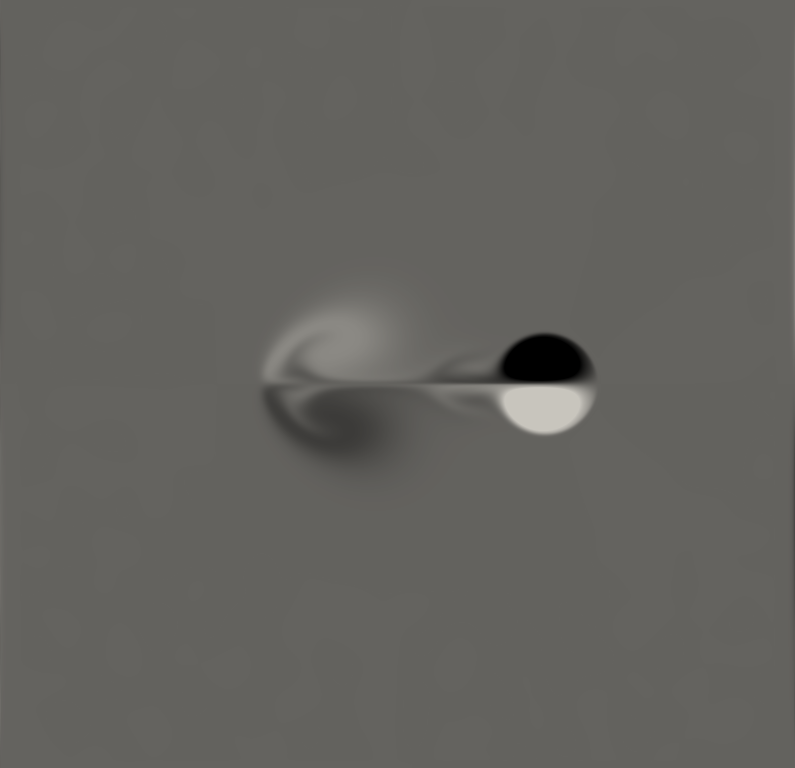}\ \includegraphics[width=.24\textwidth]{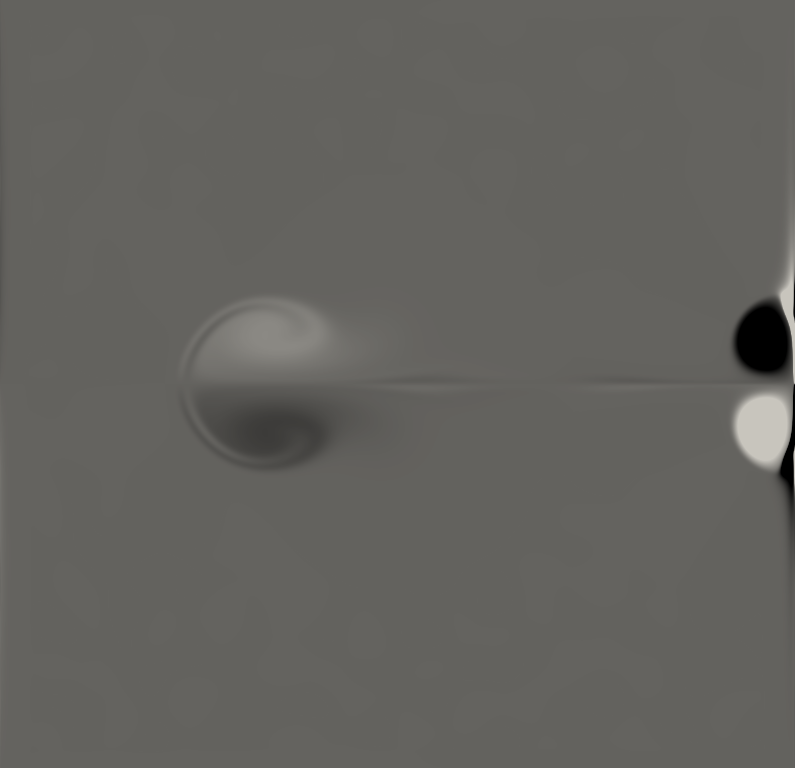}\ \includegraphics[width=.24\textwidth]{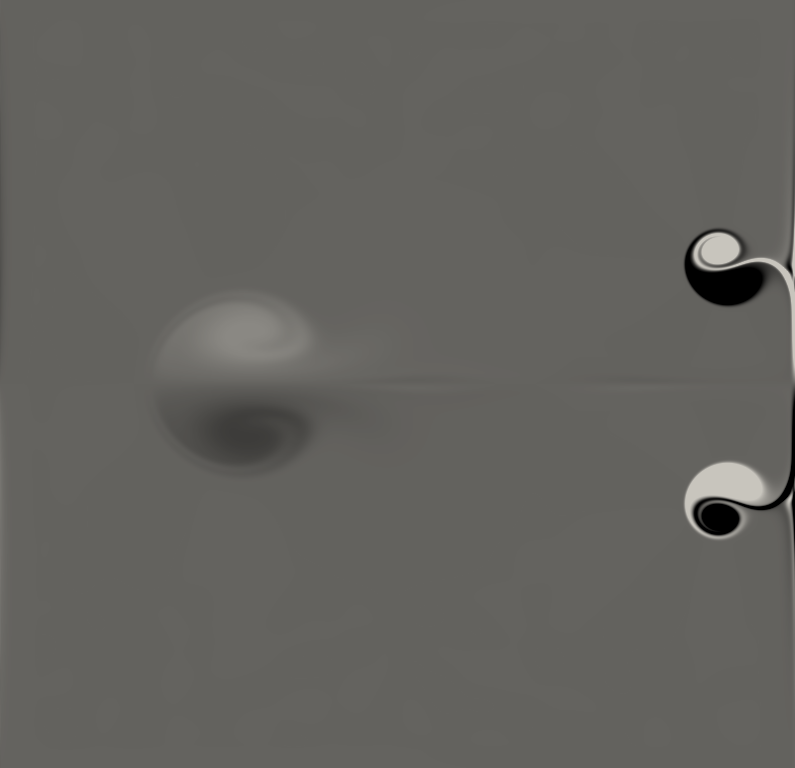}\\
\includegraphics[width=.24\textwidth]{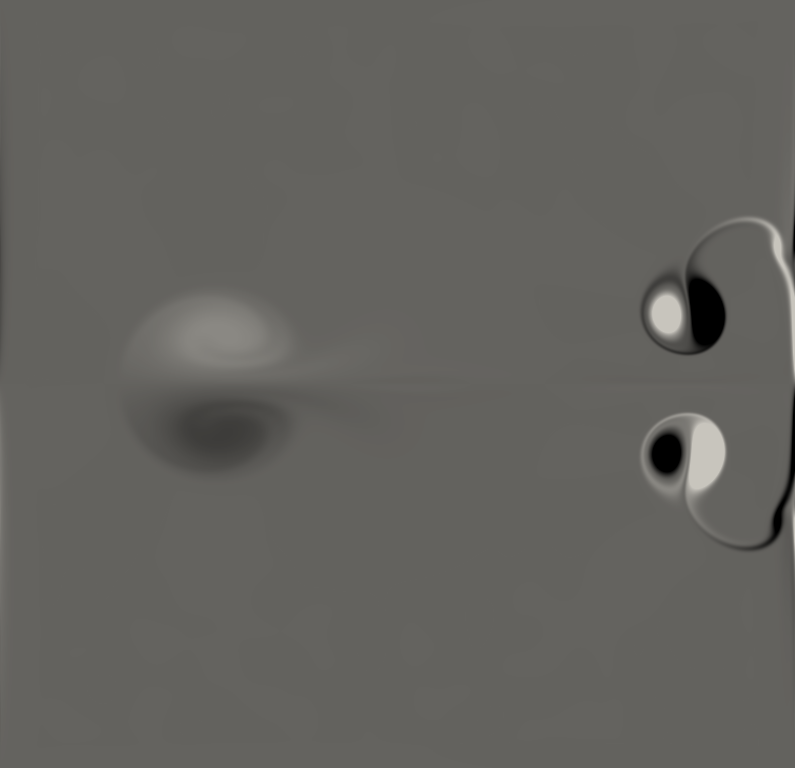}\ \includegraphics[width=.24\textwidth]{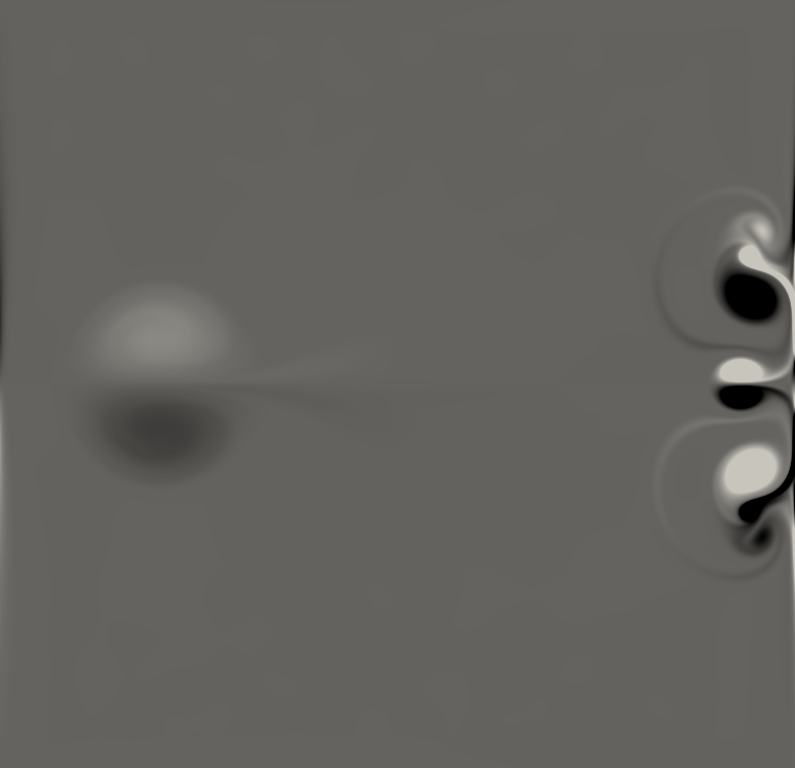}\ \includegraphics[width=.24\textwidth]{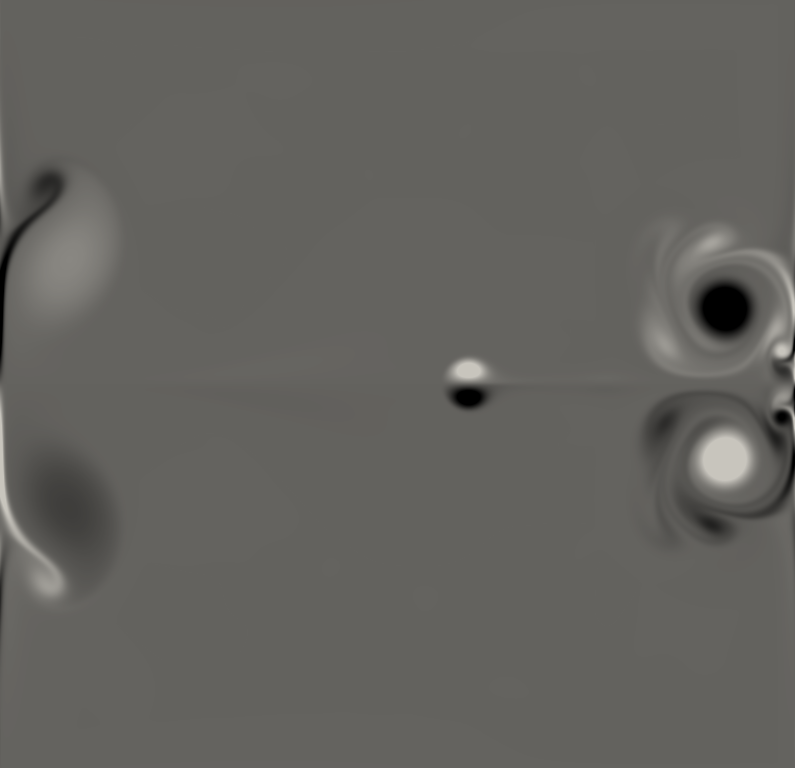}\ \includegraphics[width=.24\textwidth]{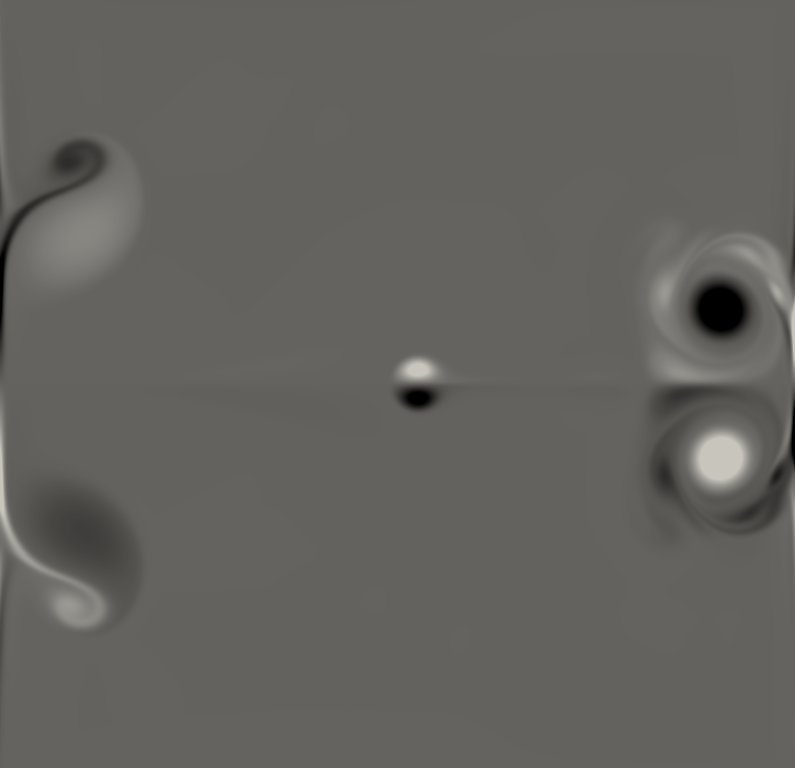}
\caption{Snapshot of the vorticity at $t=0,0.15,0.0.32,0.4,0.48,0.64,0.72.0.8$ from left to right and top to bottom. Black is for positive and white for negative vorticity.}\label{fig_vort_time}
\end{center}
\end{figure}

Under the actions of viscous forces, the dipole described by Eqs.~\eqref{eq_dipole1}
and~\eqref{eq_dipole2} develops
a net momentum in the positive $x$-direction and is self-propelled
towards the right wall.  The collision between the dipole and the wall
is characterized by a 2D turbulent dynamics where the wall acts as a
source of small-scale vortices that originate from detached boundary
layers. After the first collision the monopoles under the action of
viscosity are re-propelled against the wall. These collisions
give rise to two peaks of enstrophy (see Fig.~\ref{fig_energy_enstrophy}). The value of these local maxima
will be used for comparison with the results obtained with a spectral method in \citet{bi_clercx06}.
Several snapshot of the dynamics of the dipole--wall collision can be found on Fig.~\ref{fig_vort_time}.

In this benchmark the initial core vorticity of the monopoles is fixed to $\omega_e = 299.5286$. 
Furthermore, the Reynolds number and the monopole radius are set to $\mathrm{Re}=L U /\nu=2500$ and $r_0=0.1$. 
The positions of the monopole centers are $(x_1,y_1)=(0,0.1)$ and $(x_2,y_2)=(0,-0.1)$. 
The approach of \citet{bi_latt07} is used to set up the initial condition. 

The error on the value of the enstrophy peak is the principal quantity of interest here. It is defined 
as 
\begin{equation}
E_i=\left|\langle\Omega_{i}\rangle-\langle\Omega_{i,\mathrm{lb}}\rangle\right|/\langle\Omega_{i}\rangle, 
\end{equation}
where $i=1,2$ is the label of the enstrophy peak. The value of the enstrophy computed with the LBM (either the present model or the MRT model) is noted 
$\langle\Omega_{i,\mathrm{lb}}\rangle$. The reference value $\langle\Omega_i\rangle$ is the value found in \citet{bi_clercx06} 
and is given by $\langle\Omega_{1}\rangle=3313$ and $\langle\Omega_{2}\rangle=1418$. The convergence study is performed by keeping 
$\nu$ constant and modifying $U_\mathrm{lb}$ while varying the resolution $N$ ($\Re=U_\mathrm{lb}N/\nu$). Here $U_\mathrm{lb}=0.01 N / 125$ with $N=125,250,500,1000$.
This rescaling of the velocity has as an effect to reduce the compressibility errors from the simulation (see \citet{bi_latt-thesis}).
\begin{figure}
\begin{center}
\includegraphics[width=.45\textwidth]{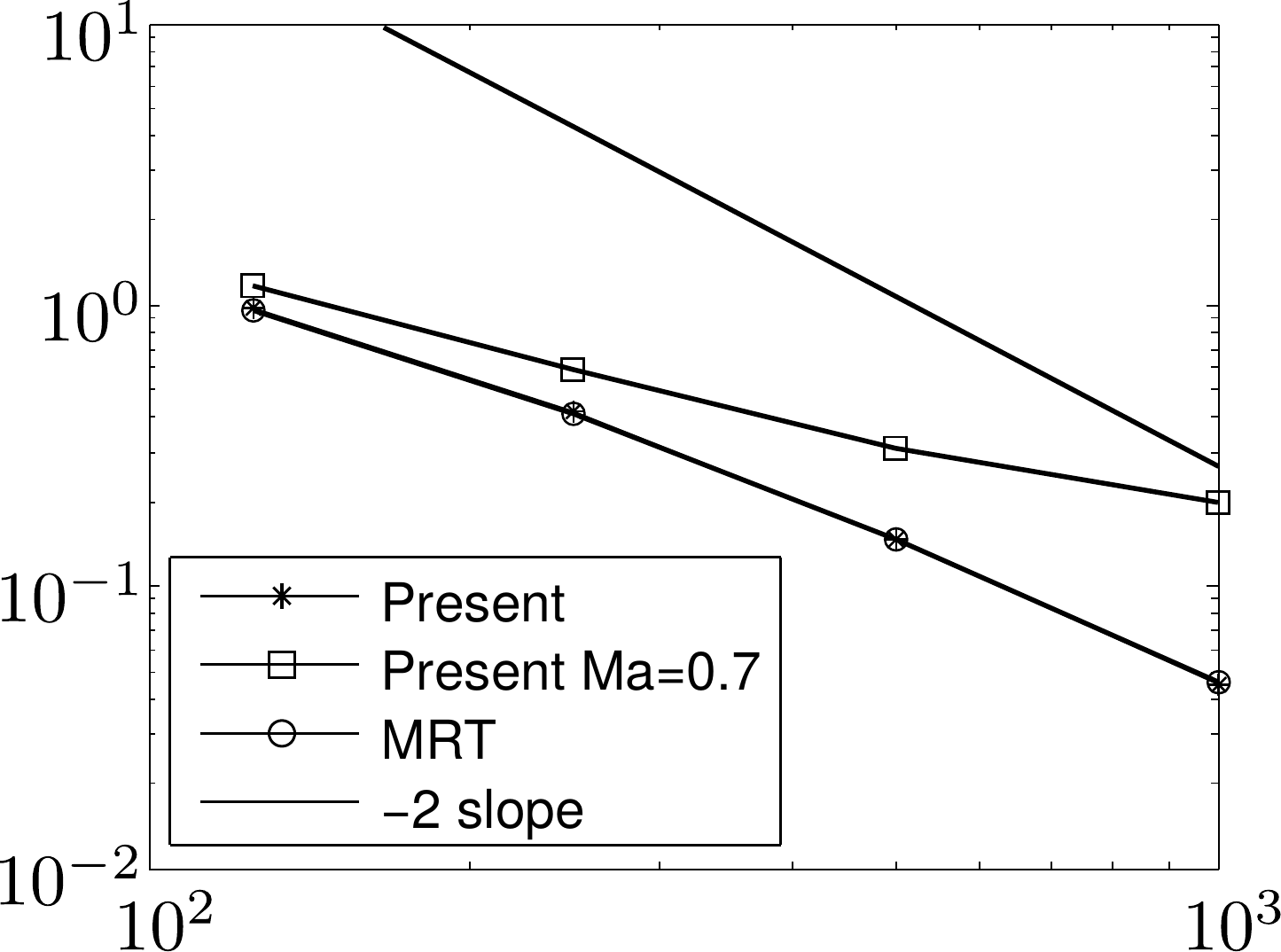}
\includegraphics[width=.45\textwidth]{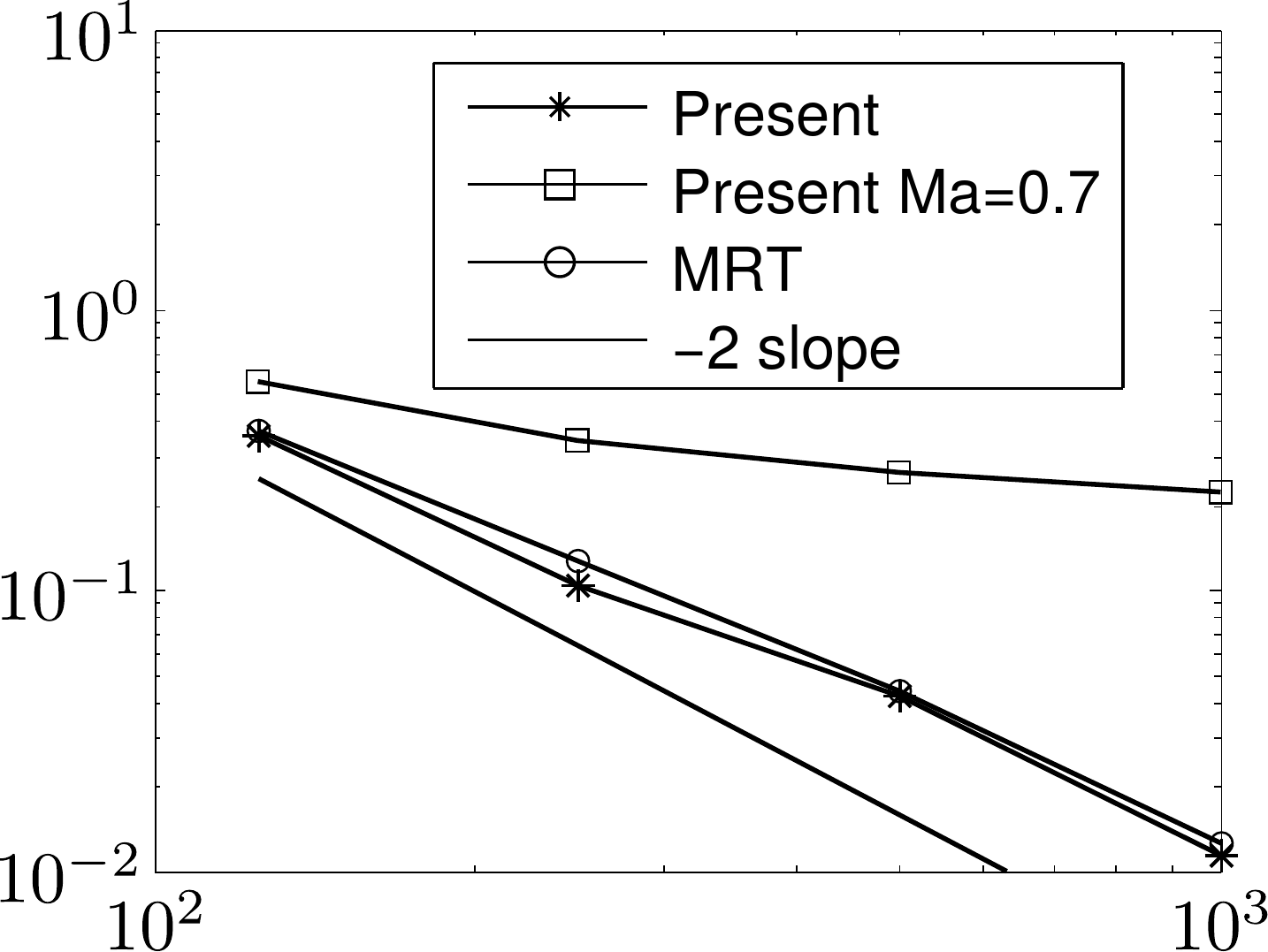}
\caption{Numerical accuracy in the 2D dipole-wall collision flow for the two enstrophy peaks, (left) $\omega_1=3313$, and (right) $\omega_2=1418$. 
The error curves for the enstrophy peak obtained with the present scheme and the MRT scheme.}\label{fig:accuracy-dipole}
\end{center}
\end{figure}
As shown by Figs.~\ref{fig:accuracy-dipole} the difference in accuracy between methods is not dramatically different. 
The differences appear when one pushes the numerical scheme to more challenging Reynolds and Mach numbers.

In order to test the numerical stability and the ability to go to ``higher'' Mach numbers (but still lower than one) we also 
simulated the dipole at a maximal $\Ma$ number of 0.7 (corresponding to a characteristic velocity of $U_\mathrm{lb}=0.032$) and $\Re=2500$. 
At such high Mach number the MRT model was numerically unstable. The maximal stable reachable Mach number 
was of 0.46 (corresponding to characteristic velocity of $U_\mathrm{lb}=0.02$).
For this test the Mach number is kept constant and therefore one modifies the viscosity (in order to keep $\Re$ constant).
By increasing the the resolution we do not remove the compressibility error terms (as discussed in \citet{bi_latt-thesis}). 
This explains the first order decrease of the error observed in Fig.~\ref{fig:accuracy-dipole} and the lower accuracy of the results.

We notice that in this case the accuracy is much lower since the compressibility effects are much higher.
Nevertheless the stability of the present model is highly enhanced with respect to the MRT model. Since such a ``high'' Mach number
was not achievable with the MRT model.

\subsection{Turbulent jet}

\begin{figure}
\begin{center}
\includegraphics[width=0.65\linewidth]{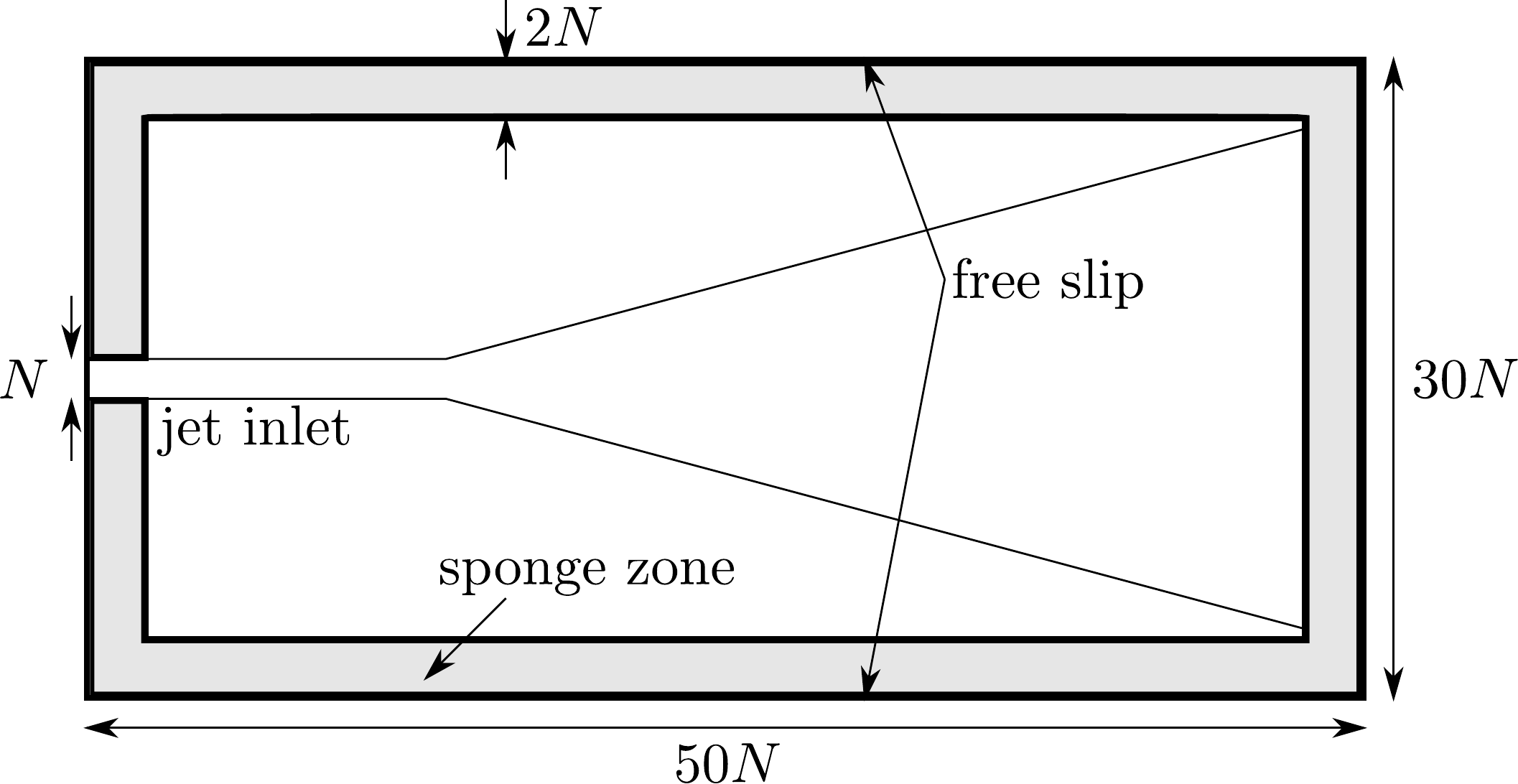}
\caption{The turbulent jet computational domain.}\label{fig_jet}
\end{center}
\end{figure}

In this section we will perform the simulation of a turbulent round jet (see \citet{bi_pope05}) at $\Re=N U/\nu=6000$ with $N$ and $U$ being the diameter and the 
speed of the 
jet respectively. The aim will be to recover correctly the self similar behavior and the correct energy spectrum, and also 
correct pressure spectrum. The computational domain is depicted on Fig.~\ref{fig_jet}. The domain size in units
of the diameter of the jet was chosen to be of $[50,30,30]\times N$. In order to avoid as much as possible 
acoustic reflexions the sponge zones proposed by \citet{bi_xu13} were added in the domain. Furthermore
a vortex ring (see \citet{bi_bogey03} for example) is added at one diameter from the inlet to help the onset of the 
instability and allow for the development of turbulence in the flow. The Mach number of the flow
is set to 0.4. The value is chosen to be rather large in order to really challenge the numerical accuracy and 
stability of the models.

The quantities of interest are defined as follows. The velocity, $\uu(x,r,\theta,t)$, is given in cylindrical coordinates centered around the center of the jet.
The mean axial velocity field at the center of the flow is given by
\begin{equation}
u_c(x)\equiv \left\langle u_x(x,0,0,t)\right\rangle,
\end{equation}
where $\langle \cdot\rangle$ is the time average.
The jet's half width, $r_{1/2}(x)$ is defined such that
\begin{equation}
 \left\langle u_x(x,r_{1/2}(x),0,t)\right\rangle=\frac{1}{2}u_c(x).
\end{equation}
We will also study the Reynolds stresses $\left\langle u'_\alpha u'_\beta\right\rangle$, where 
\begin{equation}
 \uu'=\uu-\langle \uu\rangle.
\end{equation}

The simulation is performed with a very low resolution of $N=10$ points in the diameter of the jet
and with no explicit turbulence model for the case of the present model. For the MRT case a Smagorinsky model was needed 
(see \citet{bi_krafczyk03,bi_malaspinas12}) in order to obtain stable results. 
The fact that no explicit turbulence model is needed for our novel scheme 
seems to imply that the regularization operation has the effect of an implicit turbulence model
and would deserve a more in-depth analysis.

\begin{figure}
\begin{center}
\includegraphics[width=0.45\linewidth]{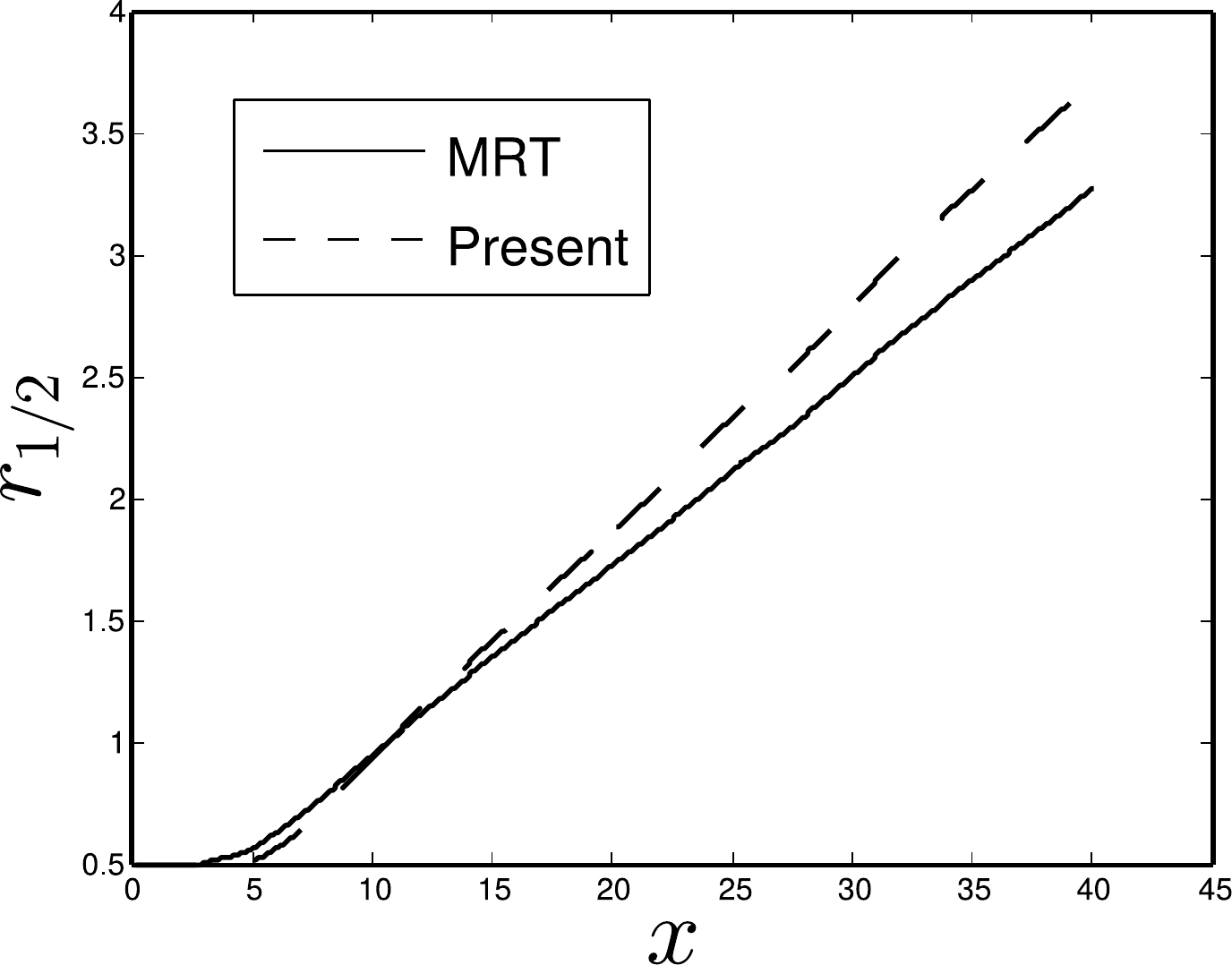}
\caption{The jet's half width with respect to the position for the MRT and the present model. The spread rate is of respectively
of $Sr_\mathrm{MRT}=0.074$ and $Sr_\mathrm{Present}=0.093$ for the MRT and present model.}\label{fig_spread_rate}
\end{center}
\end{figure}

Fig.~\ref{fig_spread_rate} depicts $r_{1/2}(x)$ from which one can compute the spread rate $Sr=\mathrm{d}r_{1/2}/\mathrm{d}x$ of 
the jet for the MRT and the present model. Although both models exhibit a self-similar behavior since there is a linear 
growth of $r_{1/2}$, the value of the spread rates are significantly different. One has respectively 
$Sr_\mathrm{MRT}=0.078$ and $Sr_\mathrm{Present}=0.093$. The expected value of the spread rate is
of roughly 0.1 (see \citet{bi_pope05}). Therefore the present model seems to provide a more accurate
result than the MRT model.

\begin{figure}
\begin{center}
\includegraphics[width=0.45\linewidth]{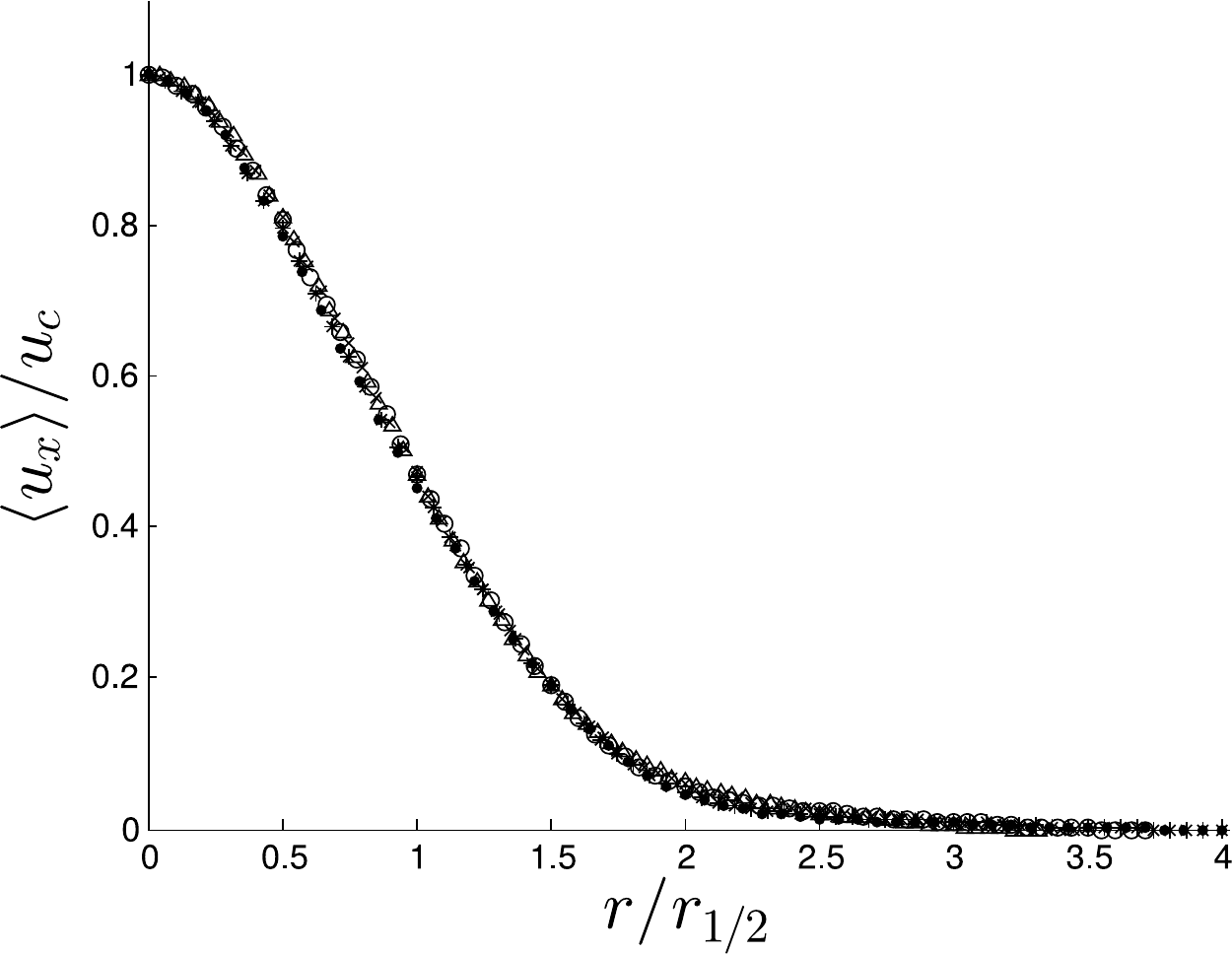}\includegraphics[width=0.45\linewidth]{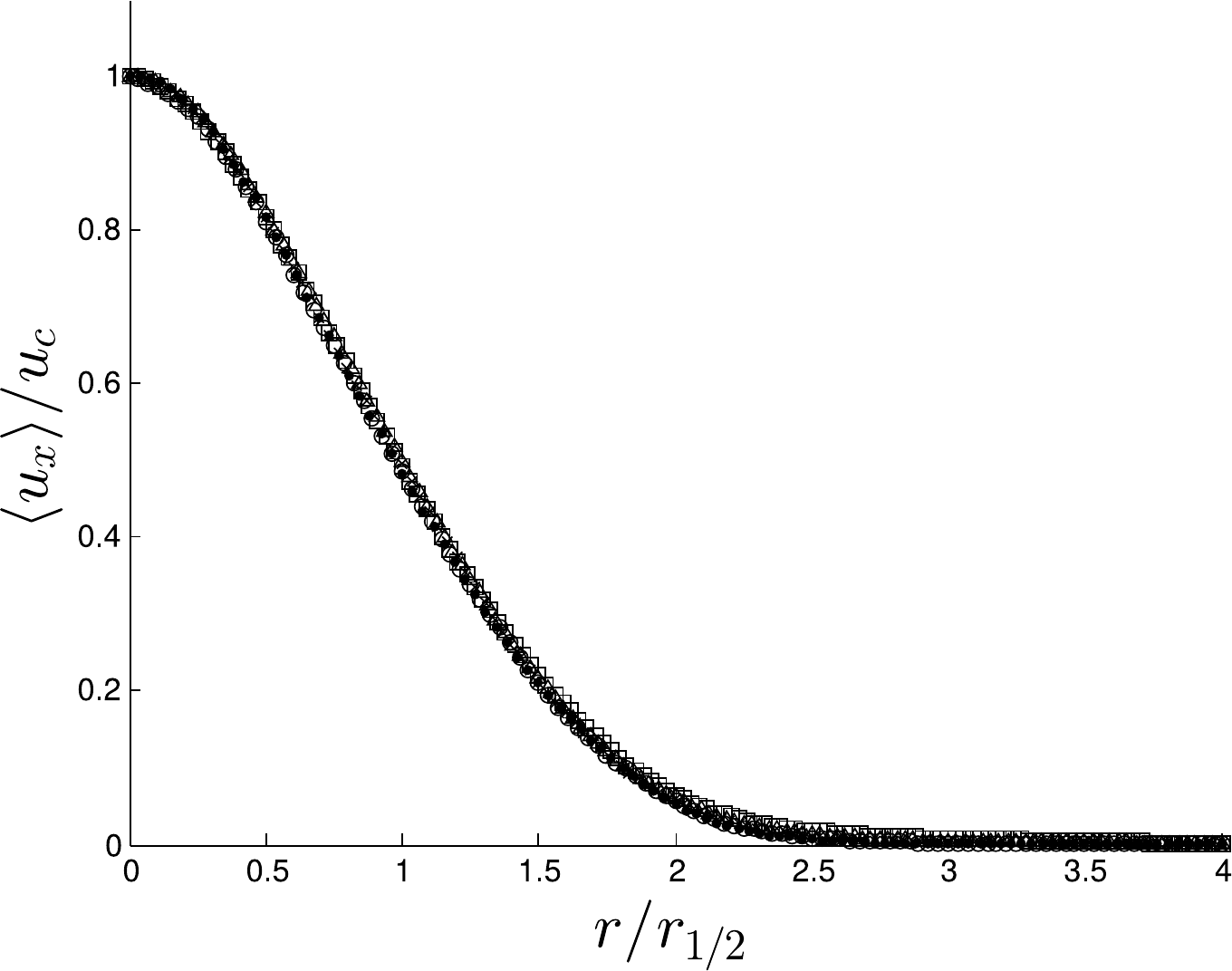}
\caption{Non-dimensional velocity profile with respect to the rescaled position for the MRT (left) and the present (right) models. 
Five different $x$ positions in part of the flow where the fluid is in a turbulent regime are depicted.}\label{fig_avg_vel}
\end{center}
\end{figure}

\begin{figure}
\begin{center}
\includegraphics[width=0.45\linewidth]{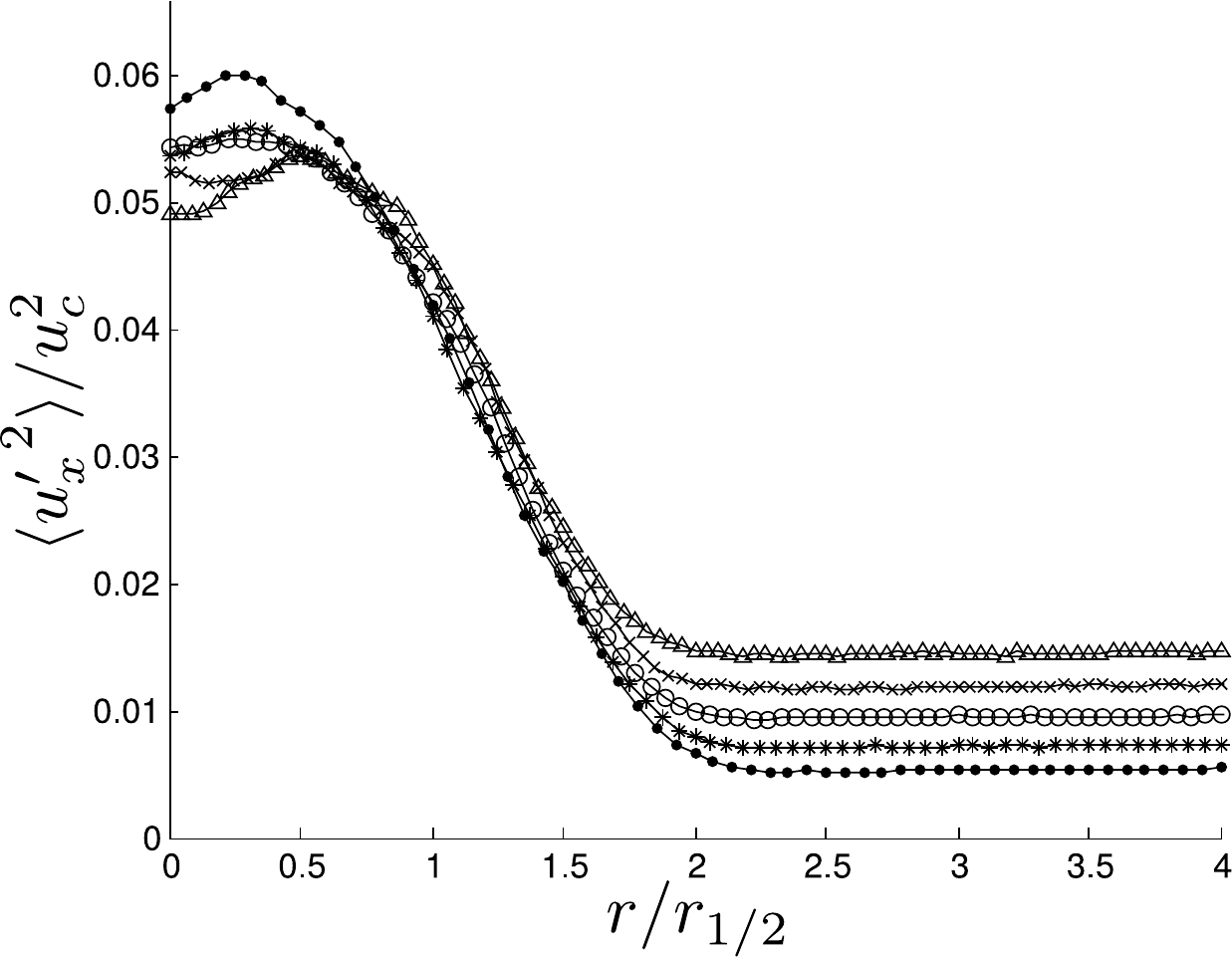}\includegraphics[width=0.45\linewidth]{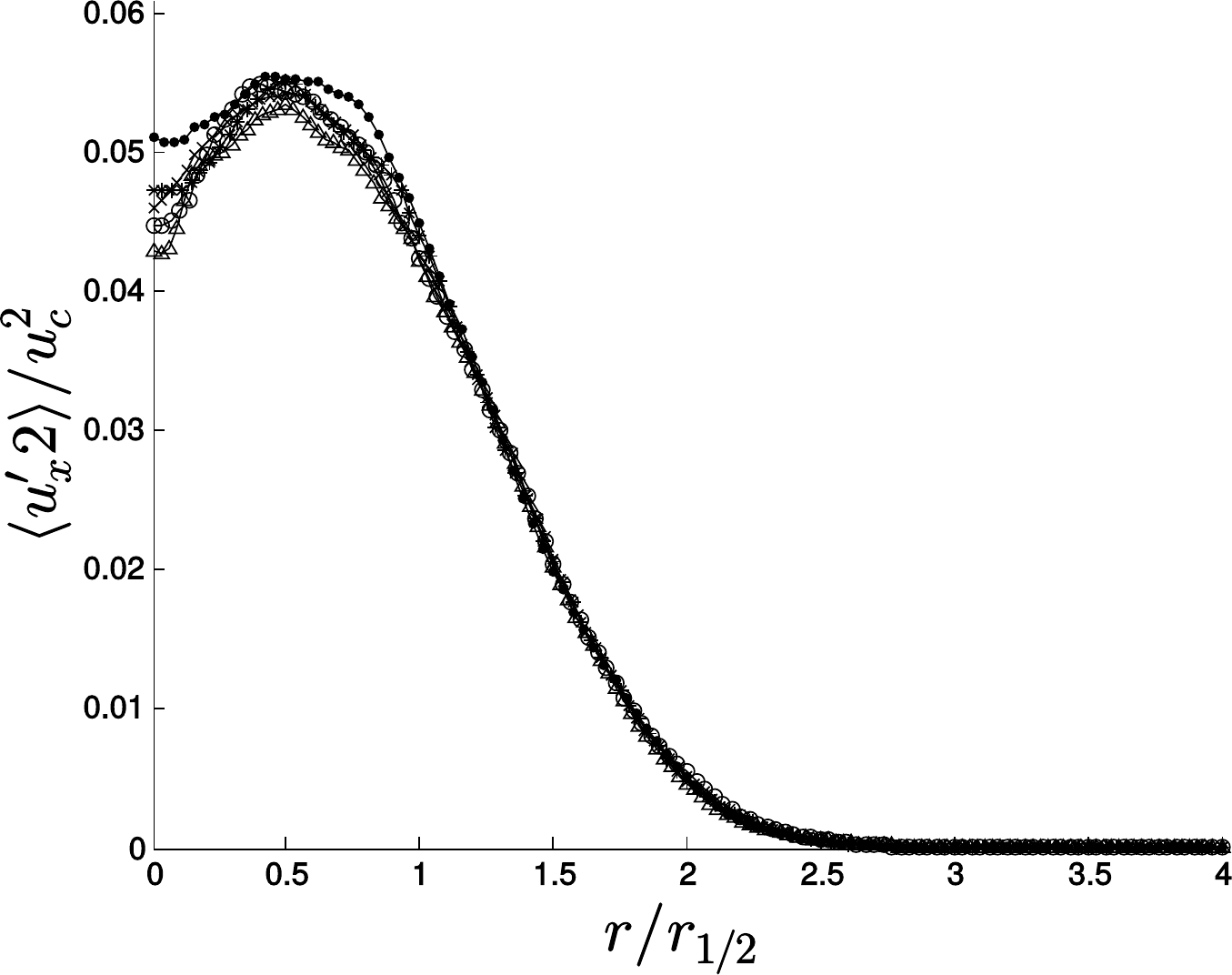}
\caption{Non-dimensional Reynolds stress component $\langle {u'_x}^2\rangle/u_c^2$ with respect to the rescaled position for the MRT (left) and the present (right) models. 
Five different $x$ positions in part of the flow where the fluid is in a turbulent regime are depicted.}\label{fig_u2}
\end{center}
\end{figure}

\begin{figure}
\begin{center}
\includegraphics[width=0.45\linewidth]{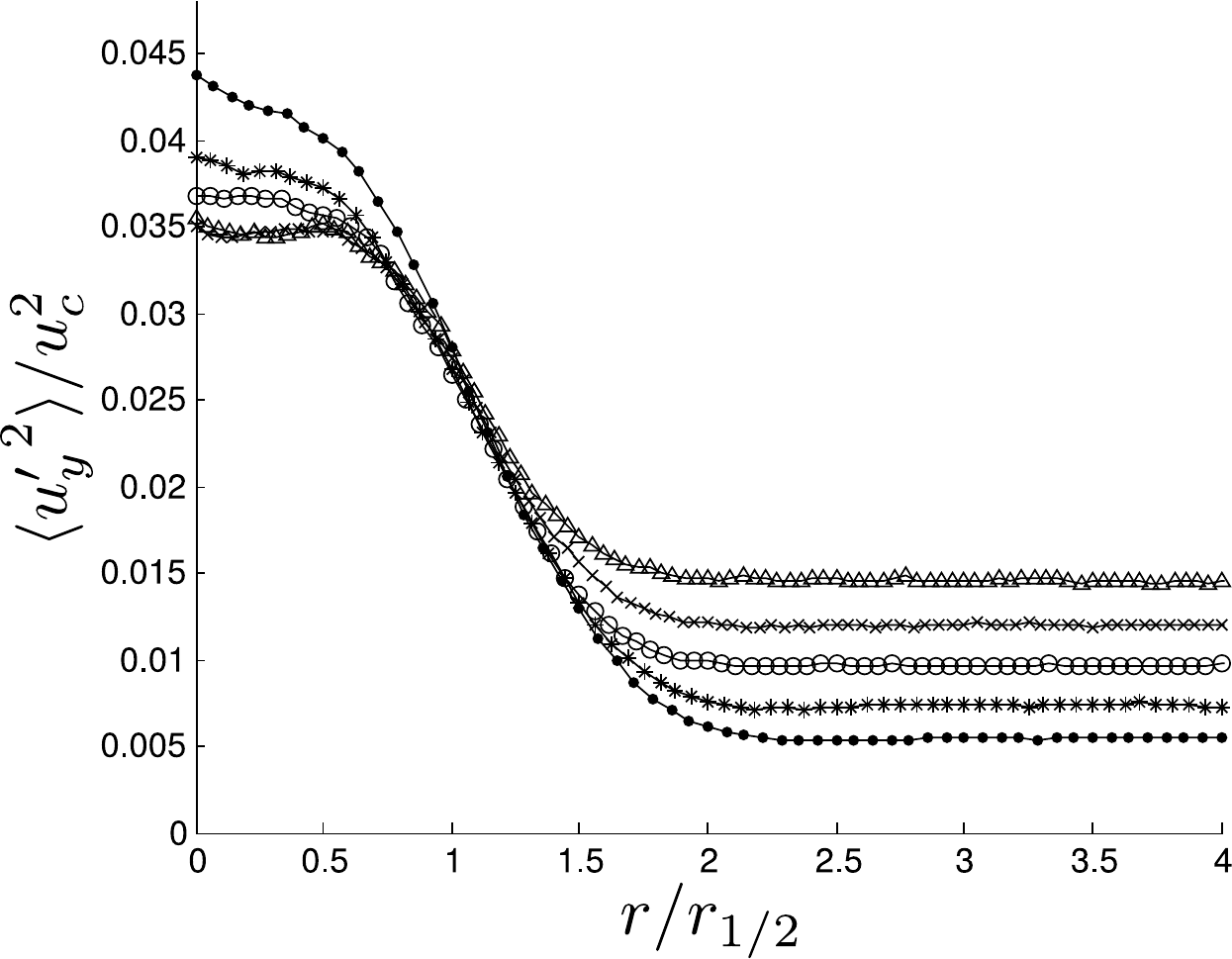}\includegraphics[width=0.45\linewidth]{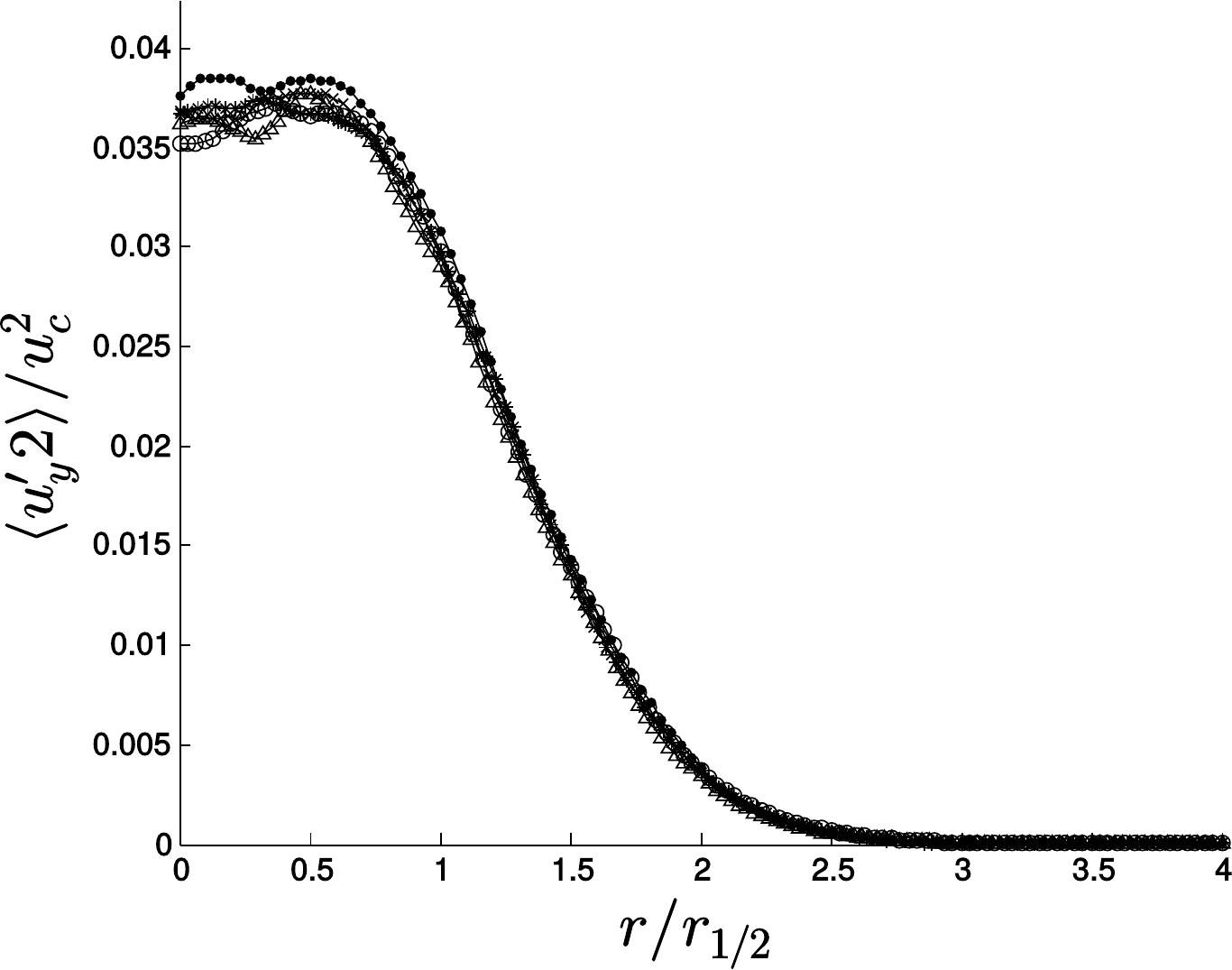}
\caption{Non-dimensional Reynolds stress component $\langle {u'_y}^2\rangle/u_c^2$ with respect to the rescaled position for the MRT (left) and the present (right) models. 
Five different $x$ positions in part of the flow where the fluid is in a turbulent regime are depicted.}\label{fig_v2}
\end{center}
\end{figure}

\begin{figure}
\begin{center}
\includegraphics[width=0.45\linewidth]{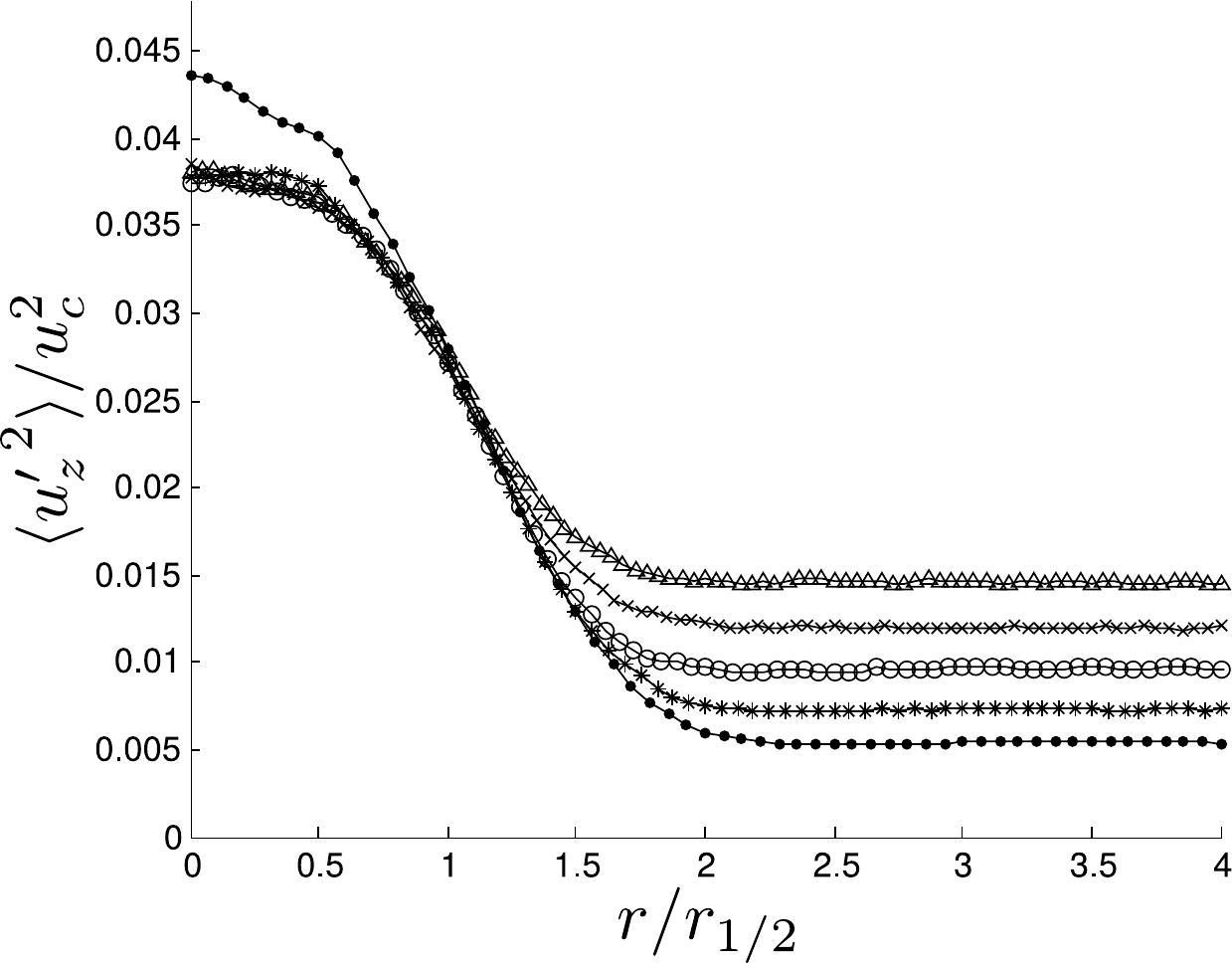}\includegraphics[width=0.45\linewidth]{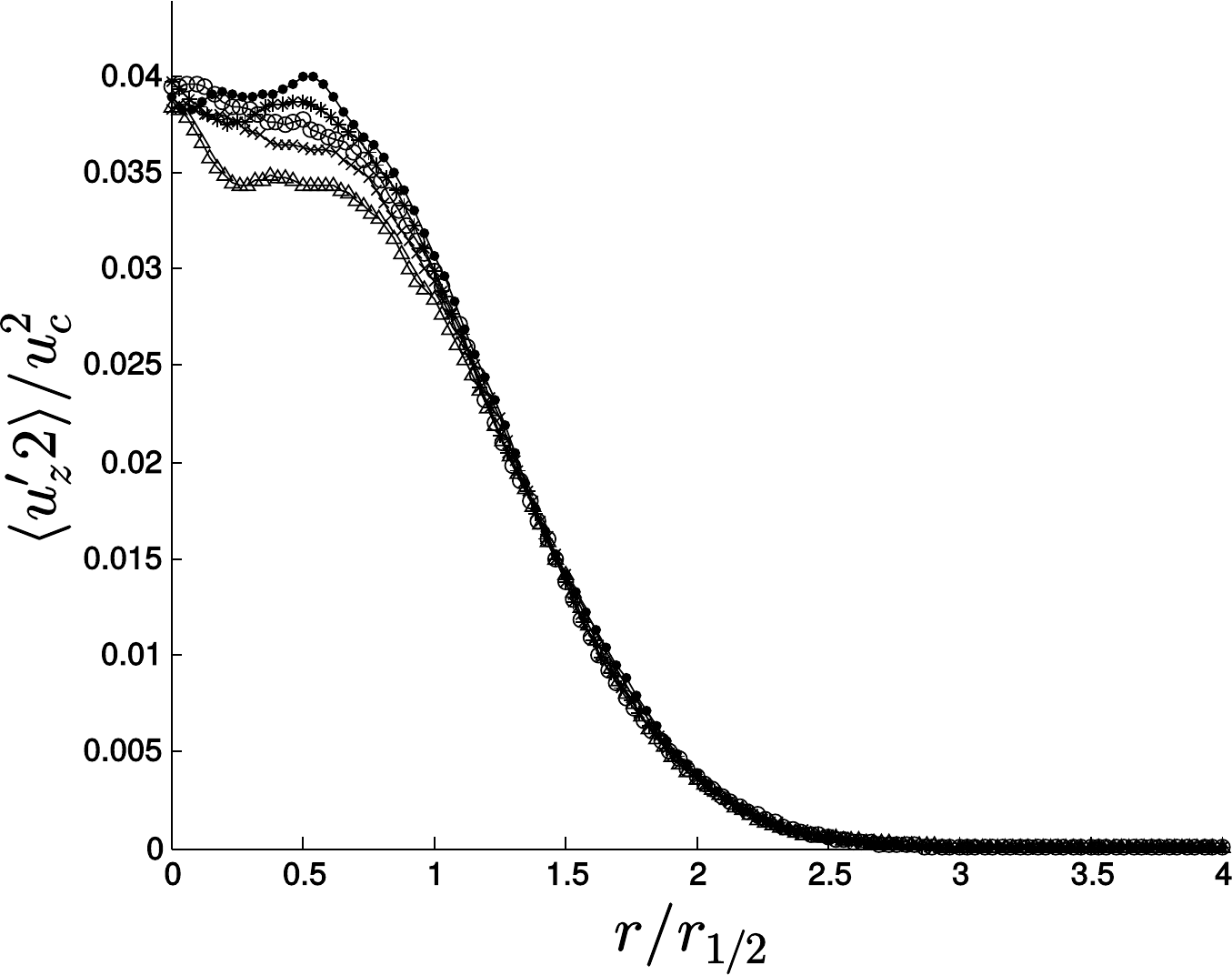}
\caption{Non-dimensional Reynolds stress component $\langle {u'_z}^2\rangle/u_c^2$ with respect to the rescaled position for the MRT (left) and the present (right) models. 
Five different $x$ positions in part of the flow where the fluid is in a turbulent regime are depicted.}\label{fig_w2}
\end{center}
\end{figure}

\begin{figure}
\begin{center}
\includegraphics[width=\linewidth]{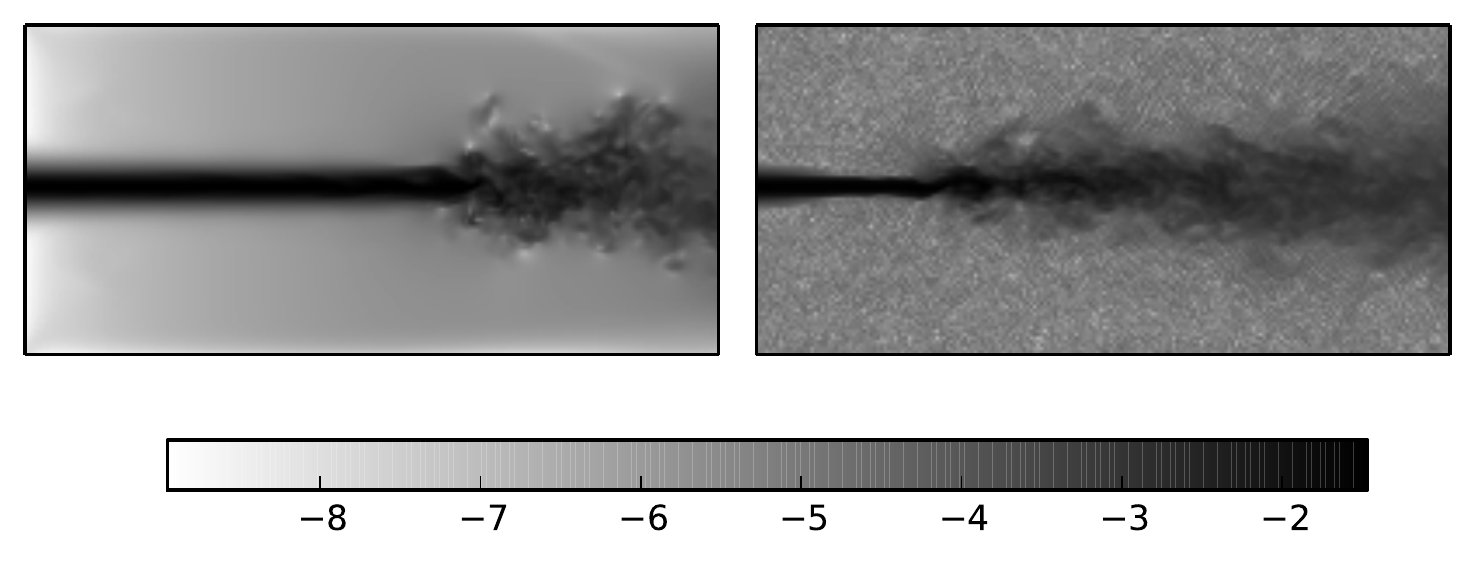}
\caption{Instantaneous velocity norm for the turbulent jet for the present model (left) and the MRT model (right) in logscale. The sponge zone region is removed from these pictures.}\label{fig_vel_jet}
\end{center}
\end{figure}

\begin{figure}
\begin{center}
\includegraphics[width=\linewidth]{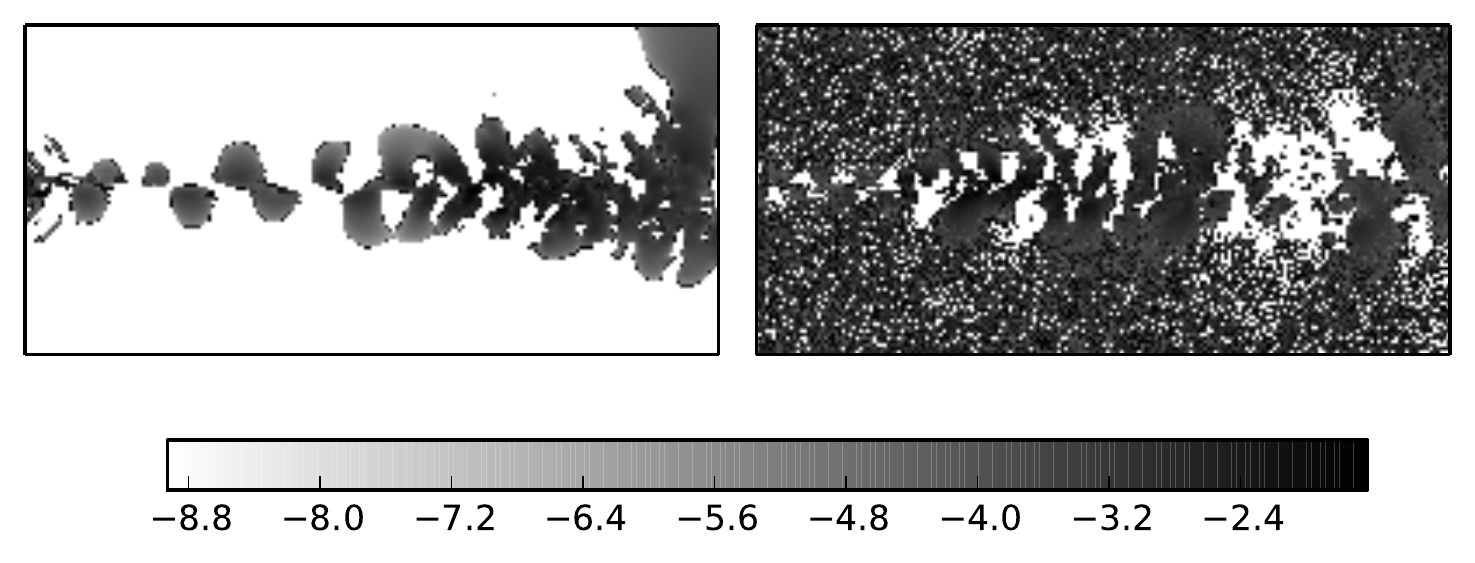}
\caption{Pressure fluctuations for the turbulent jet for the present model (left) and the MRT model (right) in logscale. The colorbar scale has been reduced to allow seeing the small acoustic perturbations. 
The sponge zone region is removed from these pictures.}\label{fig_acoustic_jet}
\end{center}
\end{figure}

As shown in Fig.~\ref{fig_avg_vel} the self-similar behavior is observed
for both models as for five different positions in the direction of the jet, the normalized average velocity profiles
overlap for the MRT and for the present model.
For the Reynolds stresses ($\langle {u'_x}^2\rangle$, $\langle {u'_y}^2\rangle$, and $\langle {u'_z}^2\rangle$ respectively) 
which are depicted on Figs.~\ref{fig_u2}-\ref{fig_w2} one can notice that the self similar behavior 
is shown for the present model. For the MRT model while close to the jet center the 
results seem self-similar (and also are coherent with what is observed with the present model), one can see that when going to $r/r_{1/2} \gtrsim 1.5$ the Reynolds stresses are not overlapping anymore
and even worse, they are not converging towards zero as they should. This behavior can be explained by looking at the instantaneous 
velocity field at a given time. As one can see from Figs.~\ref{fig_vel_jet}, and~\ref{fig_acoustic_jet} 
(which represent respectively instantaneous snapshots of the velocity norm and of the pressure fields) the 
results obtained with the present model are far less noisy and no spurious modes can be observed.
The only ``spurious'' modes present in the present regularized model
are due to the vortex ring used to trigger faster the transition to turbulence as seen on Fig.~\ref{fig_acoustic_jet}.
For the MRT spurious modes can be observed in the pressure field 
although this model is expected (see \citet{bi_lallemand00,bi_xu11} for example) to dissipate the pressure waves 
at a higher rate than for BGK models.


\begin{figure}
\begin{center}
\includegraphics[width=0.45\linewidth]{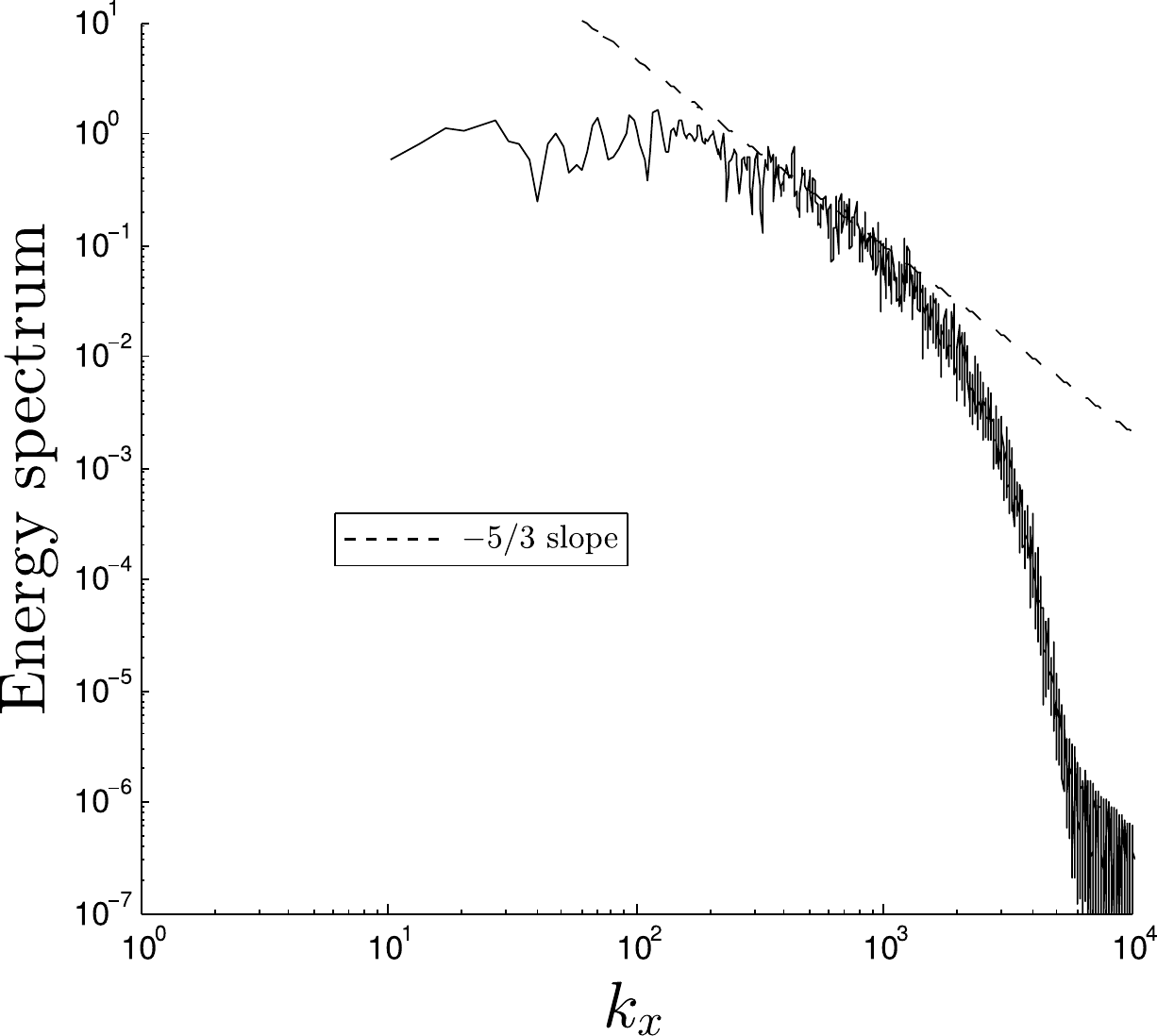}\includegraphics[width=0.45\linewidth]{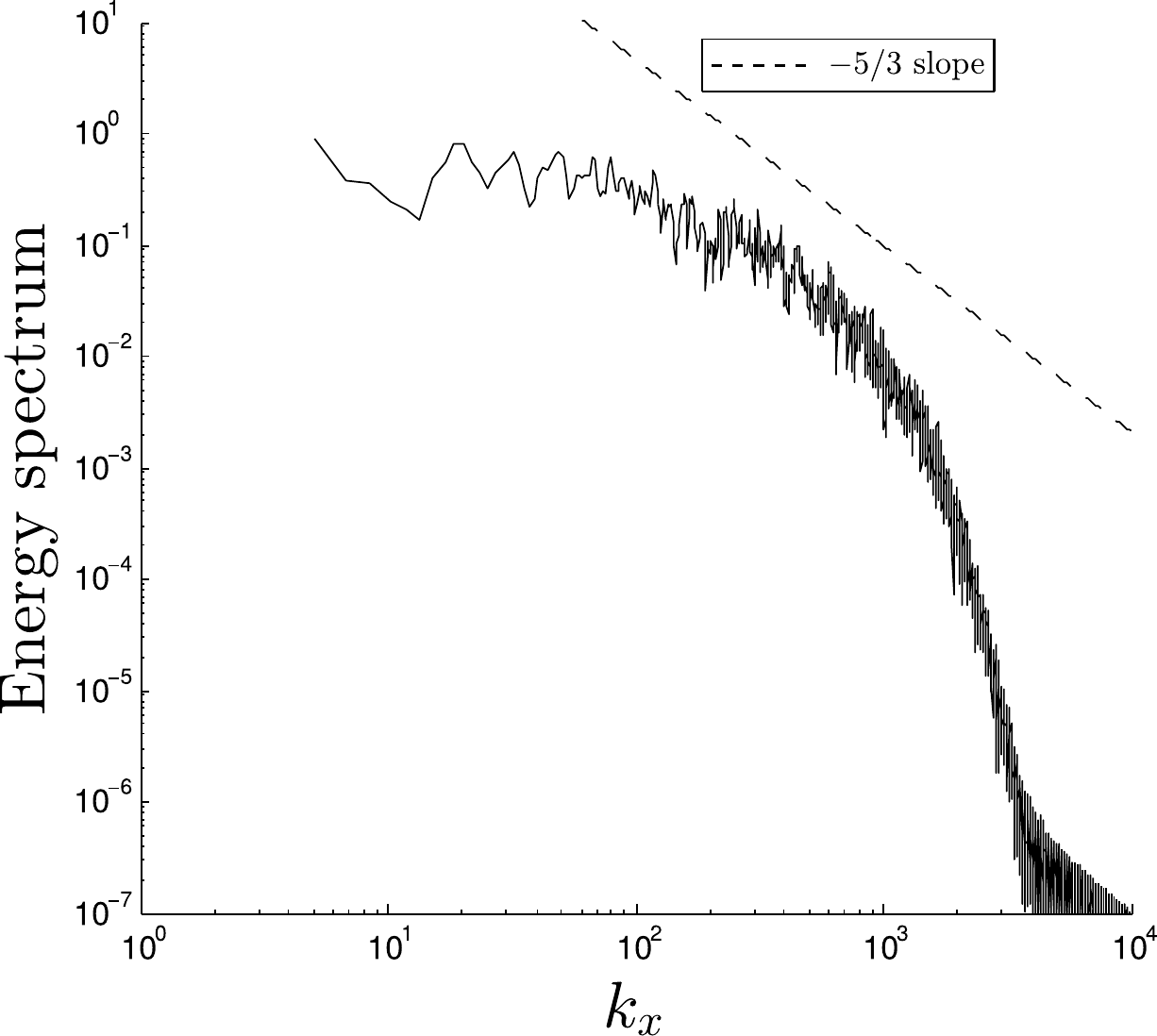}
\caption{Energy power spectrum with respect to $k_x$ for the MRT model (left) and the present model (right).}\label{fig_energy_spectrum}
\end{center}
\end{figure}

\begin{figure}
\begin{center}
\includegraphics[width=0.45\linewidth]{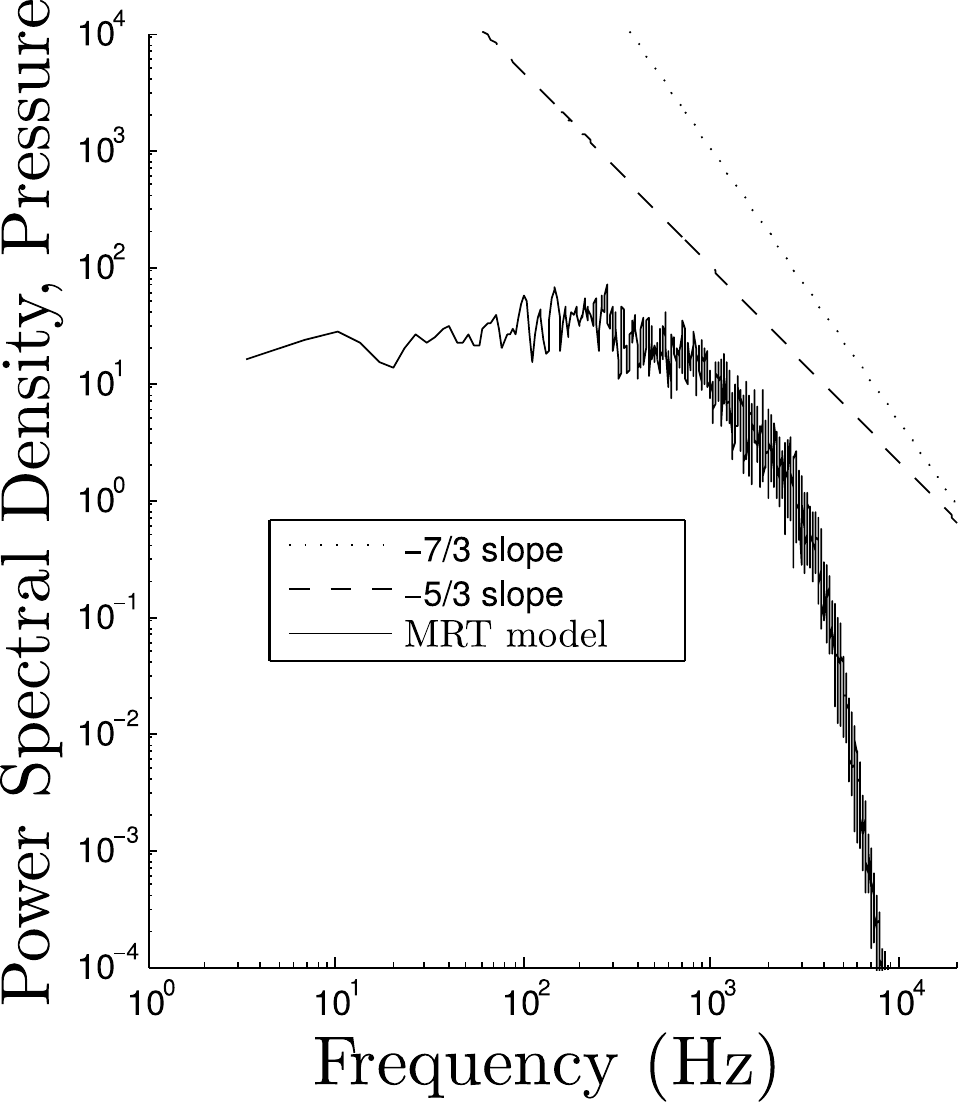}\includegraphics[width=0.45\linewidth]{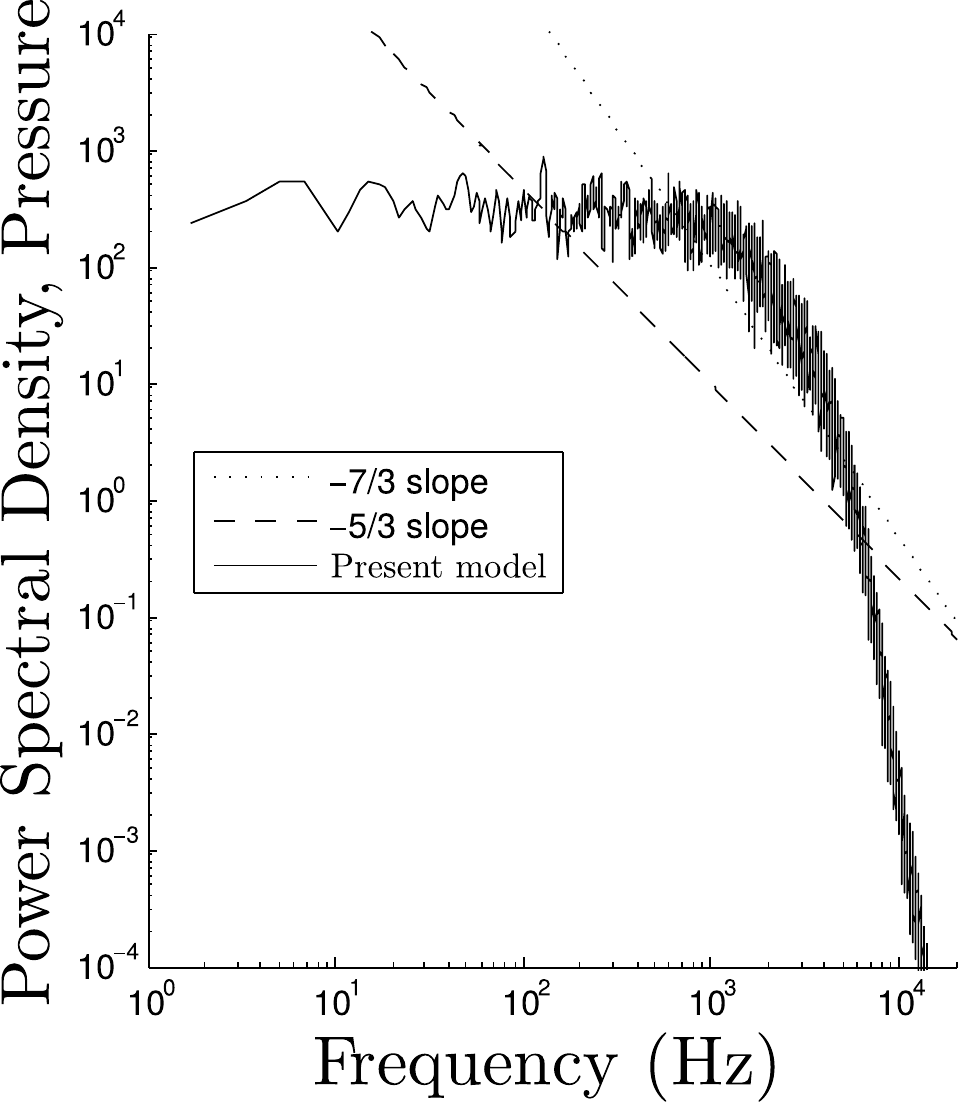}
\caption{Pressure power spectrum with respect to the frequency for the MRT model (left) and the present model (right).}\label{fig_press}
\end{center}
\end{figure}

Finally we also computed the energy and pressure power spectrum. One can see that for both the MRT and present model
the $-5/3$ slope in the inertial range is recovered for the energy spectrum (see Fig.~\ref{fig_energy_spectrum}). For the pressure spectrum one expects to 
find a $-7/3$ slope in the inertial range as shown in \citet{bi_george84}. While for the present model the pressure spectrum 
slope is correct, in the MRT case the slope of the pressure is closer to $-5/3$ (see Fig.~\ref{fig_press}). 
This difference indicates that our new model represents the dynamics of the flow with a much greater accuracy.

\section{Conclusion}\label{sec_concl}

In this paper we demonstrated the existence of a recursive formula that allows the reconstruction of the non-equilibrium moments of the 
Boltzmann-BGK equation at any order, only by knowing the lower order moments. 
This property allows to propose a regularization procedure for the BGK lattice Boltzmann method that is increasing 
the overall accuracy of the method and removing the majority of the visible spurious modes present in 
standard BGK and MRT models. This is shown by two benchmarks: the wall--dipole interaction (2D case) and
the turbulent jet (3D case).  Although in the 2D case the increase in accuracy of the model 
is not spectacular, the stability is greatly improved. For the 3D case a great improvement is shown.
Not only the Reynolds stresses are found to be more accurately represented (with respect to the MRT model) but the pressure spectrum 
has the expected behavior.

Finally one can notice that the model acts like an implicit large eddy simulation model, since the turbulent 
behavior of the jet is reproduced accurately without any need to use an explicit turbulence model.
Its relative simplicity and low cost of implementation make it very appealing
for high Reynolds number and moderately high Mach numbers (of a maximal value of roughly 0.5). 

An interesting optimization of the present scheme could be achieved by using a D3Q19/D3Q15 quadrature
instead of the D3Q27 used here. To do so the recursive formula used throughout this paper should be generalized 
to alternative basis vectors that are not Hermite polynomials.

Finally the present approach may also be extensible for 
higher order quadratures (and higher order physics where one includes temperature for example) 
like thermal and compressible flows
where one could reduce dramatically the memory needs of such cases (which is excessive 
for the moment since one needs D3Q121 quadratures). It could also provide a way to deal with boundary conditions for these kind of models, 
since with a very limited amount of information, one can reconstruct 
the complete populations with an accuracy consistent with the model.

\section*{Acknowledgments}

I would like to acknowledge Pierre Sagaut,  Bastien Chopard, Jonas Latt, Federico Brogi, and Christophe Coreixas for the enlightening discussions
and also Andreas Malaspinas for a critical proof-reading. I would like to thankfully acknowledge the support of the Swiss National Science Foundation SNF (Award PA00P2\_145364).

\begin{appendix}
\section{Computation of the recursive relation for off-equilibrium Hermite coefficients}\label{app_ce_rec}
The aim of this section is to prove Eq.~\eqref{eq_rel_a1} that we recall here
\begin{equation*}
 a^{(n)}_{1,\al_1...\al_n}=a^{(n-1)}_{1,\al_1...\al_{n-1}}u_{\al_n}+\left(u_{\al_1}...u_{\al_{n-2}}a^{(2)}_{1,\al_{n-1}\al_n}+\hbox{perm}(\al_n)\right).
\end{equation*}
To prove it we need three relations. The first is the recursive relation of the 
equilibrium distribution Hermite coefficients (see \citet{bi_shan06,bi_malaspinas-thesis})
\begin{equation}
a_{0,\al_1,...,\al_n}^{(n)}=a_{0,\al_1,...,\al_{n-1}}^{(n-1)}u_{\al_n}\hbox{, and } a_0^{(0)}=\rho. \label{eq_rec}
\end{equation}
The second is that the order zero Chapman--Enskog expansion of the continuous Boltzmann--BGK equation leads 
to the Euler equations
\begin{align}
 &\p_t \rho+\uN\cdot(\rho\uu)=0,\\
 &\rho\p_t \uu+\rho\uu\cdot\uN\uu=-\uN\rho.\label{eq_euler}
\end{align}
And finally the order one Chapman--Enskog expansion of the Hermite coefficients
\begin{align}
-\frac{1}{\tau}a_{1,\al_1,...,\al_n}^{(n)}&=\p_t a_{0,\al_1,...,\al_n}^{(n)}+\p_{\al_{n+1}} a^{(n+1)}_{0,\al_1,...,\al_{n+1}}\nonumber\\
&\quad\quad+\left(\p_{\al_1} a^{(n-1)}_{0,\al_2,...,\al_{n}}+\hbox{perm}\right),\label{eq_rec_ind}
\end{align}
where ``perm'' stands for all the cyclic permutation of indexes $\al_1,...,\al_{n-1}$.

Replacing $n$ by $n-1$ in this last equation and multiplying the result by $u_\al$ one gets
\begin{align}
-\frac{1}{\tau}u_{\al_n}a_{1,\al_1,...,\al_{n-1}}^{(n-1)}&=
u_{\al_n}\p_t a_{0,\al_1,...,\al_{n-1}}^{(n-1)}+u_{\al_n}\p_{\al_{n+1}} a^{(n)}_{0,\al_1,...,\al_{n-1},\al_{n+1}}\nonumber\\
&\quad\quad+u_{\al_n}\left(\p_{\al_1} a^{(n-2)}_{0,\al_2,...,\al_{n-2}}+\hbox{perm}\right).
\end{align}
Using the chain rule one can rewrite this equation as
\begin{align}
-\frac{1}{\tau}u_{\al_n}a_{1,\al_1,...,\al_{n-1}}^{(n-1)}&=
\underbrace{\p_t a_{0,\al_1,...,\al_{n}}^{(n)}+\p_{\al_{n+1}} a^{(n+1)}_{0,\al_1,...,\al_{n+1}}+\left(\p_{\al_1} a^{(n-1)}_{0,\al_2,...,\al_{n}}+\hbox{perm}\right)}_{(i)}\nonumber\\
&\quad\underbrace{-a^{(n-1)}_{0,\al_1,...,\al_{n-1}}\p_t u_{\al_n}-a^{(n)}_{0,\al_1,...,\al_{n-1},\al_{n+1}}\p_{\al_{n+1}} u_{\al_n}}_{(ii)}\nonumber\\
&\quad\underbrace{-\p_{\al_n}a^{(n-1)}_{0,\al_1,...,\al_{n-1}}-\left(a^{(n-2)}_{0,\al_2,...,\al_{n-2}}\p_{\al_1} u_{\al_n}+\hbox{perm}(\al_n)\right)}_{(iii)},
\end{align}
where perm$(\al_n)$ is the cyclic permutation of all the indexes not equal to $\al_n$ (the $\al_n$ index never changes position).
The $(i)$ part of the above equation is equal to (see Eq.~\eqref{eq_rec_ind})
\begin{equation}
 (i)=-\frac{1}{\tau}a_{1,\al_1,...,\al_n}^{(n)}.
\end{equation}
The $(ii)$ can be rewritten
\begin{align}
 (ii)&=-a_{0,\al_1,...,\al_{n-1}}^{(n-1)}\left(\p_t u_{\al_n}+u_{\al_{n+1}}\p_{\al_{n+1}}u_{\al_n}\right)\nonumber\\
 &=u_{\al_1}\cdots u_{\al_{n-1}}\p_{\al_n}\rho,
\end{align}
where in the first equation we used Eq.~\eqref{eq_rec} and in the second we used Eq.~\eqref{eq_euler}.
Finally the $(iii)$ part reads
\begin{align}
 (iii)&=-2\left(a_{0,\al_1,...,\al_{n-2}}^{(n-2)}S_{\al_{n-1}\al_n}+\hbox{perm}(\al_n)\right)-u_{\al_1}\cdots u_{\al_{n-1}}\p_{\al_n}\rho,\nonumber\\
 &=\frac{1}{\rho\tau}\left(a_{0,\al_1,...,\al_{n-2}}^{(n-2)}a^{(2)}_{1,\al_{n-1}\al_n}+\hbox{perm}(\al_n)\right)-u_{\al_1}\cdots u_{\al_{n-1}}\p_{\al_n}\rho,
\end{align}
where we used the chain rule in the first equation and Eq.~\eqref{eq_a1_2-classical} in the second equation.

Finally adding $(i)$, $(ii)$, and $(iii)$ one obtains
\begin{equation*}
 a^{(n)}_{1,\al_1...\al_n}=a^{(n-1)}_{1,\al_1...\al_{n-1}}u_{\al_n}+\left(u_{\al_1}...u_{\al_{n-2}}a^{(2)}_{1,\al_{n-1}\al_n}+\hbox{perm}(\al_n)\right).
\end{equation*}

\section{The higher order off-equilibrium Hermite coefficients and implementation formulas}\label{app_hermite}
In two dimensions the off-equilibrium Hermite coefficients $\ua^{(n)}_1$ are given by
\begin{align}
a^{(3)}_{1,xyy} &= u_x a^{(2)}_{1,yy}+2 u_y a^{(2)}_{1,xy}+\underbrace{\tau\rho u_y^3\p_y u_x}_{\ast},\\
a^{(3)}_{1,xxy} &= 2 u_x a^{(2)}_{1,xy}+u_y a^{(2)}_{1,xx}+\underbrace{\tau\rho u_x^3\p_x u_y}_{\ast},\\
a^{(4)}_{1,xxyy} &= u_ya^{(3)}_{xxy}+u_x^2a^{(2)}_{1,yy}+2u_x u_ya^{(2)}_{1,xy}+\underbrace{\tau\rho u_xu_y\left(u_x^2\p_x u_y+u_y^2\p_y u_x\right)}_{\ast}.
\end{align}
In three dimensions the off-equilibrium coefficients at order $n=3$ read
\begin{align}
 a^{(3)}_{1,xxy} &= 2 u_x a^{(2)}_{1,xy} +u_y a^{(2)}_{1,xx}+\underbrace{\tau\rho u_x^3\p_x u_y}_{\ast},\\
 a^{(3)}_{1,xxz} &= 2 u_x a^{(2)}_{1,xz} +u_z a^{(2)}_{1,xx}+\underbrace{\tau\rho u_x^3\p_x u_z}_{\ast},\\
 a^{(3)}_{1,xyy} &= u_x a^{(2)}_{1,yy}+2 u_y a^{(2)}_{1,xy}+\underbrace{\tau\rho u_y^3\p_y u_x}_{\ast},\\
 a^{(3)}_{1,xzz} &= u_x a^{(2)}_{1,zz}+2 u_z a^{(2)}_{1,xz}+\underbrace{\tau\rho u_z^3\p_z u_x}_{\ast},\\
 a^{(3)}_{1,yzz} &= u_y a^{(2)}_{1,zz}+2  u_z a^{(2)}_{1,yz}+\underbrace{\tau\rho u_z^3\p_z u_y}_{\ast},\\
 a^{(3)}_{1,yyz} &= 2 u_y a^{(2)}_{1,yz}+ u_z a^{(2)}_{1,yy}+\underbrace{\tau\rho u_y^3\p_y u_z}_{\ast},\\
 a^{(3)}_{1,xyz} &= u_x a^{(2)}_{1,yz}+ u_y a^{(2)}_{1,xz}+ u_z a^{(2)}_{1,xy},
 \end{align}
 while the $n=4$ are given by
 \begin{align}
 a^{(4)}_{1,xxyy} &= u_x^2 a^{(2)}_{1,yy}+2 u_x u_y a^{(2)}_{1,xy}+u_y a^{(3)}_{1,xxy}\nonumber\\
		  &\quad\quad+\underbrace{\rho \tau u_x u_y\left(2 u_y^2\p_y u_x+u_x^2\p_x u_y \right)}_{\ast},\\
 a^{(4)}_{1,xxzz} &= u_x^2 a^{(2)}_{1,zz}+2 u_x u_z a^{(2)}_{1,xz}+u_z a^{(3)}_{1,xxz}\nonumber\\
		  &\quad\quad +\underbrace{\rho \tau u_x u_z\left(u_x^2\p_x u_z +2 u_z^2\p_z u_x\right)}_{\ast},\\
 a^{(4)}_{1,yyzz} &= u_y^2 a^{(2)}_{1,zz}+2 u_y u_z a^{(2)}_{1,yz}+u_z a^{(3)}_{1,yyz}\nonumber\\
		  &\quad\quad +\underbrace{\rho \tau u_y u_z\left(u_y^2\p_y u_z +2 u_z^2\p_z u_y\right)}_{\ast},\\
 a^{(4)}_{1,xyzz} &= u_x u_y a^{(2)}_{1,zz}+u_x u_z a^{(2)}_{1,yz}+u_y u_z a^{(2)}_{1,xz}+u_z a^{(3)}_{1,xyz}\nonumber\\
		  &\quad\quad +\underbrace{\rho \tau u_z^3\left(u_x \p_z u_y +u_y \p_z u_x \right)}_{\ast},\\
 a^{(4)}_{1,xyyz} &= 2 u_x u_y a^{(2)}_{1,yz}+u_y^2 a^{(2)}_{1,xz}+u_z a^{(3)}_{1,xyy}\nonumber\\
		  &\quad\quad +\underbrace{\rho\tau u_x u_y^3\p_y u_z}_{\ast} ,\\
 a^{(4)}_{1,xxyz} &= u_x^2 a^{(2)}_{1,yz}+2 u_x u_y a^{(2)}_{1,xz}+u_z a^{(3)}_{1,xxy}\nonumber\\
		  &\quad\quad +\underbrace{\rho \tau u_y u_x^3 \p_x u_z}_{\ast}.
\end{align}
while the $n=5$ and $n=6$ are are found to be
 \begin{align}		  
 a^{(5)}_{1,xxyzz} &= u_x^2 u_y a^{(2)}_{1,zz}+u_x^2 u_z a^{(2)}_{1,yz}+2 u_x u_y u_z a^{(2)}_{1,xz}+u_z a^{(4)}_{1,xxyz}\nonumber\\
		  &\quad\quad +\underbrace{\rho \tau u_x u_z\left(u_x^2 u_y \p_x u_z +2 u_y u_z^2\p_z u_x +u_x u_z^2\p_z u_y\right)}_{\ast},\\
 a^{(5)}_{1,xxyyz} &= 2 u_x^2 u_y a^{(2)}_{1,yz}+2 u_x u_y^2 a^{(2)}_{1,xz}+u_z a^{(4)}_{1,xxyy}\nonumber\\
		  &\quad\quad +\underbrace{\rho \tau u_x^2 u_y^2\left(u_x\p_x u_z +u_y\p_y u_z \right)}_{\ast},\\
 a^{(5)}_{1,xyyzz} &= u_x u_y^2 a^{(2)}_{1,zz}+2 u_x u_y u_z a^{(2)}_{1,yz}+u_y^2 u_z a^{(2)}_{1,xz}+u_z a^{(4)}_{1,xyyz}\nonumber\\
		  &\quad\quad +\underbrace{\rho \tau u_y u_z\left(u_y u_z^2\p_z u_x +u_x u_y^2 \p_y u_z +2 u_x u_z^2\p_z u_y \right)}_{\ast},\\
 a^{(6)}_{1,xxyyzz} &= u_x^2 u_y^2 a^{(2)}_{1,zz}+2 u_x^2 u_y u_z a^{(2)}_{1,yz}+2 u_x u_y^2 u_z a^{(2)}_{1,xz}+u_z a^{(5)}_{1,xxyyz}\nonumber\\
		  &\quad\quad +\underbrace{\rho \tau u_x u_y u_z\left(u_x^2 u_y\p_x u_z +2 u_y u_z^2\p_z u_x +u_x u_y^2 \p_y u_z +2 u_x u_z^2\p_z u_y \right)}_{\ast}.
\end{align}
The $\ast$ terms are the ones that are not present in the continuous case (see Eq.~\eqref{eq_rel_a1}) and are due to the 
discretization of the microscopic velocity space. One can see that as pointed out in Subsec.~\ref{sec_ce_mod} these terms 
are of order $\O(\Ma^{n+1})$ for the coefficients of order $n$, which makes them one order of magnitude smaller 
than the other composing them. Therefore they are simply ignored for the computation of the $\ua^{(n)}_1$ terms.
\end{appendix}

\bibliographystyle{plainnat}

\end{document}